\documentclass[11pt,a4paper]{article}
\pdfoutput=1

\usepackage[T1]{fontenc}
\usepackage{jheppub}
\usepackage{graphicx,epsfig,amsmath,amssymb}
\usepackage{prettyref}
\usepackage{subcaption}
\usepackage{enumitem}
\usepackage{calc}
\usepackage{mathtools}
\usepackage{float}
\usepackage{booktabs}
\usepackage[section]{placeins}
\usepackage{multirow}
\usepackage{hyperref}
\usepackage{cleveref}
\usepackage{comment}
\usepackage{soul}

\usepackage{wrapfig}

\setlist[itemize]{nosep,left=0.5\parindent .. \parindent}

\input{all.tikzdefs}

\tikzstyle{blank}=[fill=white, shape=circle, draw=white, inner sep=0.8pt]
\tikzstyle{dot}=[fill=black, shape=circle, draw=black, inner sep=0.8pt]
\tikzstyle{fat}=[fill=white, shape=circle, draw={rgb,255: red,176; green,36; blue,39}, dashed, line width=1pt, inner sep=0.8pt]

\tikzstyle{edge}=[-, draw=PropagatorColor, line width=1.2pt, line cap=rect, preaction={{draw=white,line width=2.5pt}}]
\tikzstyle{incoming edge}=[line width=1pt, line cap=rect, draw={rgb,255: red,102; green,102; blue,102}, {|-}]
\tikzstyle{outgoing edge}=[line width=1pt, line cap=rect, draw={rgb,255: red,102; green,102; blue,102}, ->]
\tikzstyle{external edge}=[line width=1pt, line cap=rect, draw={rgb,255: red,102; green,102; blue,102}, -]
\tikzstyle{top}=[-, draw=TopPropagatorColor, line width=1.9pt, line cap=rect, preaction={{draw=white,line width=2.5pt}}]
\tikzstyle{edge dot1}=[-, postaction=decorate, decoration={markings,mark=at position .50 with {\node[style=dot]{};}}]
\tikzstyle{edge dot2}=[-, postaction=decorate, decoration={markings,mark=between positions 0.33 and 0.67 step 0.33 with {\node[style=dot]{};}}]
\tikzstyle{edge dot3}=[-, postaction=decorate, decoration={markings,mark=between positions 0.25 and 0.76 step 0.25 with {\node[style=dot]{};}}]
\tikzstyle{edge dot4}=[-, postaction=decorate, decoration={markings,mark=between positions 0.20 and 0.81 step 0.20 with {\node[style=dot]{};}}]
\tikzstyle{dot1}=[-, draw=none, postaction=decorate, decoration={markings,mark=at position .50 with {\node[style=dot]{};}}]
\tikzstyle{dot2}=[-, draw=none, postaction=decorate, decoration={markings,mark=between positions 0.33 and 0.67 step 0.33 with {\node[style=dot]{};}}]
\tikzstyle{dot3}=[-, draw=none, postaction=decorate, decoration={markings,mark=between positions 0.25 and 0.76 step 0.25 with {\node[style=dot]{};}}]
\tikzstyle{dot4}=[-, draw=none, postaction=decorate, decoration={markings,mark=between positions 0.20 and 0.81 step 0.20 with {\node[style=dot]{};}}]
\tikzstyle{incoming}=[line width=1pt, line cap=rect, draw=LegColor, {|-}]
\tikzstyle{outgoing}=[line width=1pt, line cap=rect, draw=LegColor, ->]
\tikzstyle{outgoing top}=[line width=1pt, line cap=rect, draw=TopLegColor, ->]
\tikzstyle{outgoing higgs}=[line width=1pt, line cap=rect, draw=HiggsLegColor, ->]
\tikzstyle{edge}=[-, draw={rgb,255: red,176; green,36; blue,39}, line width=1pt, preaction={{draw=white,line width=2pt}}, line cap=rect]
\tikzstyle{massive edge}=[-, draw={rgb,255: red,133; green,119; blue,181}, line width=2.0pt, preaction={{draw=white,line width=2.5pt}}, line cap=rect]
\tikzstyle{cut edge}=[-, draw={rgb,255: red,64; green,64; blue,64}, line width=0.5pt, densely dashed, line cap=rect]
\tikzstyle{xcut edge}=[-, draw={rgb,255: red,64; green,64; blue,64}, line width=0.5pt, densely dashed, line cap=rect, postaction={decorate, decoration={markings,mark=at position .50 with {\node[cross out,solid,draw=white,line width=2pt,inner sep=1.4pt,transform shape] {};}}}, postaction={decorate, decoration={markings,mark=at position .50 with {\node[cross out,solid,draw={rgb,255: red,176; green,36; blue,39},line width=1pt,inner sep=1.8pt,transform shape] {};}}}]
\tikzstyle{xcut edge 1/3}=[-, draw={rgb,255: red,64; green,64; blue,64}, line width=0.5pt, densely dashed, line cap=rect, postaction={decorate, decoration={markings,mark=at position .33 with {\node[cross out,solid,draw=white,line width=2pt,inner sep=1.4pt,transform shape] {};}}}, postaction={decorate, decoration={markings,mark=at position .33 with {\node[cross out,solid,draw={rgb,255: red,176; green,36; blue,39},line width=1pt,inner sep=1.8pt,transform shape] {};}}}]
\tikzstyle{cut}=[-, draw={rgb,255: red,61; green,171; blue,83}, line width=0.7pt, dotted, line cap=rect]
\tikzstyle{photon}=[-, draw={rgb,255: red,176; green,36; blue,39}, line width=1pt, preaction={{draw=white,line width=2pt}}, line cap=rect, decorate, decoration=snake]
\tikzstyle{gluon}=[-, draw={rgb,255: red,176; green,36; blue,39}, line width=1pt, preaction={{draw=white,line width=2pt}}, line cap=rect, decorate, decoration={coil,aspect=1.4,segment length=2.5mm}]
\tikzstyle{gluoncoil}=[-, decorate, decoration={coil,aspect=1.4,segment length=2.5mm}]
\tikzstyle{fermion}=[-, draw={rgb,255: red,176; green,36; blue,39}, line width=1pt, preaction={{draw=white,line width=2pt}}, line cap=rect, postaction=decorate, decoration={markings,mark=at position .60 with {\arrow{stealth[round]}}}]
\tikzstyle{ghost}=[-, style=fermion, line width=1pt, line cap=round, dash pattern={on 0pt off 3\pgflinewidth}]
\tikzstyle{scalar}=[-, line width=1pt, densely dashed, draw={rgb,255: red,102; green,102; blue,102}]
\tikzstyle{fermionarrow}=[-,postaction=decorate, decoration={markings,mark=at position .60 with {\arrow{stealth[round]}}}]
\tikzstyle{arrow}=[{-{Classical TikZ Rightarrow[length=2mm,width=1.5mm]}}, draw={rgb,255: red,61; green,171; blue,83}, line width=1pt, preaction={{draw=white,line width=2pt}}, line cap=rect]
\tikzstyle{brace}=[-,draw={rgb,255: red,61; green,171; blue,83}, line width=1pt, decorate, decoration={brace,amplitude=5pt}]

\usepackage{xcolor}
\usepackage{mathtools}
\usepackage{amsmath}
\usepackage{prettyref}
\usepackage{dsfont}
\usepackage{bookmark}

\allowdisplaybreaks

\definecolor{SapGreen}{HTML}{3B7B3B}
\definecolor{PhtaloGreen}{HTML}{2b7a76}
\definecolor{EmeraldGreen}{HTML}{1ea78d}
\definecolor{EnglishRed}{HTML}{b02427}
\definecolor{IronOxideRed}{HTML}{a13634}
\definecolor{CadmiumRed}{HTML}{df2b3f}
\definecolor{Vermillion}{HTML}{e14235}

\definecolor{Black}{HTML}{000000}

\hypersetup{colorlinks=true,urlcolor=EmeraldGreen,citecolor=EmeraldGreen,linkcolor=EnglishRed}

\let\origref\ref
\newrefformat{chap}{\hyperref[#1]{Chapter~\origref*{#1}}}
\newrefformat{app}{\hyperref[#1]{Appendix~\origref*{#1}}}
\newrefformat{sec}{\hyperref[#1]{Section~\origref*{#1}}}
\newrefformat{subsec}{\hyperref[#1]{Section~\origref*{#1}}}
\newrefformat{tab}{\hyperref[#1]{Table~\origref*{#1}}}
\newrefformat{fig}{\hyperref[#1]{Figure~\origref*{#1}}}
\newrefformat{eq}{eq.\,\hyperref[#1]{(\origref*{#1})}}
\AtBeginDocument{\renewcommand*{\ref}[1]{\prettyref{#1}}}

\ifcsname pdfmdfivesum\endcsname
\else
    \ifcsname mdfivesum\endcsname
        
    \else
        \errmessage{Neither pdfmdfivesum nor mdfivesum commands are present. Fail}
    \fi
\fi

\newcommand{\noun}[1]{\textsc{#1}}
\newcommand{\code}[1]{\texttt{#1}}

\input{all.tikzdefs}

\global\long\def\d{\mathrm{d}}%
\global\long\def\V#1{\mathbf{#1}}%
\global\long\def\M#1{\mathbb{#1}}%
\global\long\def\PS{\mathrm{PS}}%
\global\long\def\sp{\!\cdot\!}%
\global\long\def\ep{\varepsilon}%
\global\long\def\fn#1#2{#1\!\left(#2\right)}%
\global\long\def\undernote#1#2{\underbrace{#1}_{\mathclap{#2}}}%

\usepackage{hyperref}
\hypersetup{pdfusetitle,colorlinks=true,urlcolor=purple,citecolor=purple,linkcolor=black}

\let\plainref\ref
\newrefformat{chap}{\hyperref[#1]{Chapter~\plainref*{#1}}}
\newrefformat{app}{\hyperref[#1]{Appendix~\plainref*{#1}}}
\newrefformat{sec}{\hyperref[#1]{Section~\plainref*{#1}}}
\newrefformat{subsec}{\hyperref[#1]{Section~\plainref*{#1}}}
\newrefformat{tab}{\hyperref[#1]{Table~\plainref*{#1}}}
\newrefformat{fig}{\hyperref[#1]{Figure~\plainref*{#1}}}
\newrefformat{eq}{eq.\,\hyperref[#1]{(\plainref*{#1})}}
\newrefformat{fun}{\hyperref[#1]{\plainref*{#1}}}
\newrefformat{lab}{\hyperref[#1]{\plainref*{#1}}}
\AtBeginDocument{\renewcommand*{\ref}[1]{\prettyref{#1}}}

\def\d{\mathrm{d}}

\title{Master integrals for semi-inclusive cuts \\of massless four-loop propagators}
\hypersetup{pdftitle={Master integrals for semi-inclusive cuts of massless 4-loop propagators}}
\author[a]{Vitaly Magerya,}
\author[b,c]{Levente Fekésházy}
 
\affiliation[a]{Theoretical Physics Department, CERN, 1211 Geneva 23, Switzerland}
\affiliation[b]{II. Institut für Theoretische Physik, Universität Hamburg,\\ Luruper Chaussee 149, 22761 Hamburg, Germany}

\affiliation[c]{
    Institute for Theoretical Physics, ELTE E\"otv\"os Lor\'and University, \\
	P\'azm\'any P\'eter s\'et\'any 1/A,
    1117,
    Budapest,
    Hungary
    }

\emailAdd{vitaly.magerya@cern.ch}
\emailAdd{levente.fekeshazy@desy.de}

\preprint{{\footnotesize\texttt{CERN-TH-2025-061, DESY-25-048}}}

\abstract{
    We present the analytic calculation of all master integrals for 3-, 4-, and 5-particle semi-inclusive cuts of four-loop massless propagators by means of differential equations.
    We fix the integration constants by reducing the semi-inclusive integrals to their fully inclusive counterparts with the Integration-By-Parts (IBP) method.
    We validate our results by calculating the next-to-next-to-next-to-leading order (N$^3$LO) semi-inclusive $e^+e^- \to X$ cross-section and integrating directly over the differential variable.
    We provide the results in a machine-readable form.
    The presented integrals are essential for the direct calculation of NNLO time-like splitting functions and N$^3$LO coefficient functions, which play a crucial role in precision QCD calculations.
}

\usepackage{tocloft}
\renewcommand\afterTocRuleSpace{}

\begin{document}

\maketitle

\section{Introduction}

Theoretical predictions for semi-inclusive hadron production at $e^+ e^-$ colliders, such as the planned FCC-ee~\cite{FCC18}, ILC~\cite{ILC13},
CLIC~\cite{CLIC18}, or CEPC~\cite{CEPC2018}, require the knowledge of perturbative QCD theory at high precision.
Central to this are photonic coefficient functions, currently known up to order $\alpha_s^2$~\cite{RN96,RN97a,RN97b}, and time-like splitting functions, currently known up to $\alpha_s^3$~\cite{MMV06,MV08,AMV12,Che+20}.

To upgrade the coefficient functions to the next order, or to re-calculate the splitting functions directly, one needs the knowledge of semi-inclusive scattering cross-section for $e^+ e^- \to q + X$ up to $\alpha_s^3$.
This cross-section can naturally be expressed in terms of semi-inclusive cuts of propagators (two-point functions).
The order $\alpha_s^3$ requires cuts of 4-loop propagators.



Master integrals for semi-inclusive cuts of 3-loop propagators have been previously calculated in~\cite{Git16}.
Semi-inclusive 2-particle cuts of 4-loop propagators correspond trivially to inclusive cuts, and their master integrals are known from~\cite{HHM08,Hei+09,LSS10}.

In this article we shall present the calculation of all master integrals for 3-, 4-, and 5-particle semi-inclusive cuts of massless 4-loop propagators, which was first developed as a part of \cite{M22}, recalculate those of 3-loop propagators, and provide machine-readable tables for both sets of results via~\cite{xcut4lfiles}.
For this calculation we shall employ the method of differential equations~\cite{Hen13,Lee15}, where the integration constants are obtained by matching with the corresponding fully inclusive integrals, as previously advocated in~\cite{GM15,Git16}.
The reduction of differential equations to $\ep$-form is automatized via \noun{Fuchsia}, see~\cite[Chap.\,8]{M22} and~\cite{GM17,GM16}. 
The required fully inclusive integrals were completed in~\cite{MP19,GMP18a,MM06}.


\section{Semi-inclusive phase-space integrals\protect\label{chap:semi-inclusive-phase-space-ints}}

A \emph{semi-inclusive cut of a propagator} in $d=4-2\ep$ dimensions,
with $L$ loop momenta, $P$ denominators, and $n$ cut momenta has
the form
\begin{equation}
    I=\int\frac{\d^{d}l_{1}}{\left(2\pi\right)^{d}}\cdots\frac{\d^{d}l_{L}}{\left(2\pi\right)^{d}}\frac{1}{D_{1}^{\nu_{1}}\cdots D_{P}^{\nu_{P}}}
    \undernote{\fn{\d\PS_{n}}q\fn{\delta}{x-2\frac{q\sp p_{n}}{q^{2}}}}{\equiv\fn{\d\PS_{n}}{q,x}},
    \label{eq:semiinclusive-int}
\end{equation}
where $q$ is the incoming momentum, $l_{i}$ are the loop momenta, $D_{i}$ are the denominators,
$\nu_i$ are their indices (exponents), and $\fn{\d\PS_{n}}q$
is the phase-space volume element,
\begin{equation}
    \d\fn{\PS_{n}}{q} \equiv
        \left(\prod_{i=1}^{n}\frac{\d^{d}p_{i}}{\left(2\pi\right)^{d}}2\pi\fn{\delta^{+}}{p_{i}^{2}}\right)\left(2\pi\right)^{d}\fn{\delta^{d}}{q-\sum_{j=1}^{n}p_{j}}.
    \label{eq:dpsn}
\end{equation}
For cuts of 4-loop propagators ($L+n=5$), there are 2-, 3-, 4-, and 5-particle cuts, for example:
\begin{equation}
    \raisebox{0.5ex}{\scalebox{0.75}{\input{fig/si/b16-line.tikz}}},\quad
    \raisebox{0.5ex}{\scalebox{0.75}{\input{fig/si/b135-line.tikz}}},\quad
    \raisebox{0.5ex}{\scalebox{0.75}{\input{fig/si/b172-line.tikz}}},\quad
    \raisebox{0.5ex}{\scalebox{0.75}{\input{fig/si/b243-line.tikz}}}.
\end{equation}
The dashed lines on these Feynman diagrams correspond to cut (on-shell) propagators; the cross marks the $x$-tagged momentum $p$ entering $\fn{\delta}{x-2q\sp p/q^2}$.

Integrals of this form arise from squared amplitudes integrated
over the phase space.
An equivalent \emph{fully inclusive integral} would be the same, but
without $\fn{\delta}{x-2\,{q\sp p_{n}}/{q^{2}}}$; this
factor makes integrals differential (i.e.\ not inclusive) in
$x$, hence ``semi-inclusive''.

\subsection{Semi-inclusive phase space}

The simplest semi-inclusive integral is the semi-inclusive phase space,
\begin{equation}
    \fn{\PS_{n}}{q,x}=\raisebox{0.5ex}{\scalebox{0.75}{\input{fig/psnx.tikz}}}=\int\fn{\d\PS_{n}}q\fn{\delta}{x-2\frac{q\sp p_{n}}{q^{2}}}.
\end{equation}
To calculate it, note that the $\delta$ function here does not
depend on $p_{i<n}$, so we can factorize the equation as
\begin{equation}
    \fn{\PS_{n}}{q,x}=\int\fn{\d\PS_{n-1}}{q-p_{n}}\frac{\d^{d-1}\vec{p}_{n}}{\left(2\pi\right)^{d-1}}\frac{1}{2\left|\vec{p}_{n}\right|}\fn{\delta}{x-2\frac{q\sp p_{n}}{q^{2}}}.
\end{equation}
The $\left(n-1\right)$-particle phase-space here can be integrated
over right away using the known
\begin{equation}
    \fn{\PS_{n}}q=\left(q^{2}\right)^{n\left(\frac{d}{2}-1\right)-\frac{d}{2}}\frac{2\pi}{\left(4\pi\right)^{\frac{d}{2}\left(n-1\right)}}\frac{\fn{\Gamma^{n}}{\frac{d}{2}-1}}{\fn{\Gamma}{\left(\frac{d}{2}-1\right)\left(n-1\right)}\fn{\Gamma}{\left(\frac{d}{2}-1\right)n}}.\label{eq:psn}
\end{equation}
Next, we can move into the frame of reference where $q=\left(q,\vec{0}\right)$,
so that $q\sp p_{n}$ becomes $q\left|\vec{p}_{n}\right|$, and the
integrand is manifestly invariant to the rotations of $\vec{p}_{n}$,
\begin{equation}
    \fn{\PS_{n}}{q,x} = \PS_{n-1}(1)\int\frac{\d^{d-1}\vec{p}_{n}}{\left(2\pi\right)^{d-1}}\frac{1}{2\left|\vec{p}_{n}\right|}\left(q^{2}-2q\left|\vec{p}_{n}\right|\right)^{\left(n-1\right)\left(\frac{d}{2}-1\right)-\frac{d}{2}}\fn{\delta}{x-2\frac{\left|\vec{p}_{n}\right|}{q}}.
\end{equation}
This allows us to integrate out the angular degrees of freedom of
$\d^{d-1}\vec{p}_{n}$, which results in the surface area of a $\left(d-2\right)$-sphere,
or $\left|\vec{p}_{n}\right|^{d-2}\Omega_{d-2}$ via \ref{eq:omega}.
Then, we are only left with the radial part of~$\d^{d-1}\vec{p}_{n}$,
\begin{equation}
    \fn{\PS_{n}}{q,x}=\PS_{n-1}(1)\int\frac{\d\left|\vec{p}_{n}\right|}{\left(2\pi\right)^{d-1}}\frac{\left|\vec{p}_{n}\right|^{d-3}\Omega_{d-2}}{2}\left(q^{2}-2q\left|\vec{p}_{n}\right|\right)^{\left(n-1\right)\left(\frac{d}{2}-1\right)-\frac{d}{2}}\undernote{\fn{\delta}{x-2\frac{\left|\vec{p}_{n}\right|}{q}}}{=\fn{\delta\,}{\left|\vec{p}_{n}\right|-\frac{1}{2}qx}\frac{q}{2}}.
\end{equation}
Finally, integrating over $\left|\vec{p}_{n}\right|$ resolves the
$\delta$ function,
\begin{equation}
    \fn{\PS_{n}}{q,x}=\PS_{n-1}(1)\frac{\left(\frac{1}{2}qx\right)^{d-3}\Omega_{d-2}}{2\left(2\pi\right)^{d-1}}\left(q^{2}\left(1-x\right)\right)^{\left(n-1\right)\left(\frac{d}{2}-1\right)-\frac{d}{2}}\frac{q}{2},
\end{equation}
or
\begin{equation}
    \fn{\PS_{n}}{q,x}=\fn{\PS_{n}}q\frac{\fn{\Gamma}{\left(\frac{d}{2}-1\right)n}}{\fn{\Gamma}{d-2}\fn{\Gamma}{\left(\frac{d}{2}-1\right)\left(n-2\right)}}x^{d-3}\left(1-x\right)^{n\left(\frac{d}{2}-1\right)-d+1}.\label{eq:psnx}
\end{equation}
A good consistency check for this result is that an explicit integration
over all $x$ gives back the fully inclusive integral:
\begin{equation}
    \int_{0}^{1}\d x\,\fn{\PS_{n}}{q,x} = \fn{\PS_{n}}{q}.
\end{equation}
Note that this integration is only
possible if the powers of both $x$ and $\left(1-x\right)$ factors
are greater than $-1$,
\begin{equation}
\left\{ \begin{array}{l}
d-3>-1\\
\left(\frac{d}{2}-1\right)n-d+1>-1
\end{array}\right.\qquad\Longleftrightarrow\qquad\left\{ \begin{array}{l}
d>2\\
n>2
\end{array}\right.,\label{eq:psnx-restriction}
\end{equation}
otherwise the integration would diverge at the boundaries.

\subsubsection*{Two-particle semi-inclusive phase space}

We would like to highlight the importance of the restriction on the $n=2$ case, since the massless two-particle phase-space is kinematically restricted through the momentum conservation.
If a particle of momentum $q$ decays into two massless particles of
momenta $p_{1}$ and $p_{2}$, then
\begin{equation}
    q=p_{1}+p_{2}\qquad\Rightarrow\qquad q^{2}=\undernote{p_{1}^{2}}{=0}+2p_{1}\sp p_{2}+\undernote{p_{2}^{2}}{=0}.
\end{equation}
From the definition of the variable $x$ in \ref{eq:semiinclusive-int},
only one value of $x$ is possible in this case:
\begin{equation}
    x=1,
\end{equation}
so the two-particle semi-inclusive phase space has distributional
nature:
\begin{equation}
    \fn{\PS_{2}}{q,x}=\fn{\PS_{2}}q\fn{\delta}{x-1}.\label{eq:dps2x}
\end{equation}
Because of this, master integrals for 2-particle semi-inclusive cuts correspond directly to 2-particle inclusive cuts.

\section{Semi-inclusive cuts of four-loop propagators}

To access $\alpha_{s}^{3}$ corrections we need all sets of cuts of
4-loop propagators: 2-particle cuts, 3-particle, 4-particle, and 5-particle
cuts. Among these, the 2-particle cuts correspond to 3-loop form-factors,
and have been completed in~\cite{HHM08,Hei+09,LSS10}.
The 5-particle cuts are purely phase-space integrals (no loop part); the calculation of inclusive version of these has been presented in~\cite{GMP18a}.
A subset of 3- and 4-particle inclusive cuts have been calculated in~\cite{Cza+15}; the full set was completed in~\cite{MP19}.
The master integrals for 4-loop propagators themselves (without cuts) are known from \cite{BC10,LSS12}; we list them in \ref{tab:vvvv}; master integrals for both inclusive and semi-inclusive cuts appear as various cuts of these integrals.

\begin{table}[t]
\begin{centering}
\scalebox{0.75}{%
\begin{tabular}{ccccccc}
\raisebox{0.5ex}{\scalebox{0.75}{\input{fig/prop/1.tikz}}} & \raisebox{0.5ex}{\scalebox{0.75}{\input{fig/prop/2.tikz}}} & \raisebox{0.5ex}{\scalebox{0.75}{\input{fig/prop/3.tikz}}} & \raisebox{0.5ex}{\scalebox{0.75}{\input{fig/prop/4.tikz}}} & \raisebox{0.5ex}{\scalebox{0.75}{\input{fig/prop/5.tikz}}} & \raisebox{0.5ex}{\scalebox{0.75}{\input{fig/prop/6.tikz}}} & \raisebox{0.5ex}{\scalebox{0.75}{\input{fig/prop/7.tikz}}}\tabularnewline
 &  &  &  &  &  & \tabularnewline
\raisebox{0.5ex}{\scalebox{0.75}{\input{fig/prop/8.tikz}}} & \raisebox{0.5ex}{\scalebox{0.75}{\input{fig/prop/9.tikz}}} & \raisebox{0.5ex}{\scalebox{0.75}{\input{fig/prop/10.tikz}}} & \raisebox{0.5ex}{\scalebox{0.75}{\input{fig/prop/11.tikz}}} & \raisebox{0.5ex}{\scalebox{0.75}{\input{fig/prop/12.tikz}}} & \raisebox{0.5ex}{\scalebox{0.75}{\input{fig/prop/13.tikz}}} & \raisebox{0.5ex}{\scalebox{0.75}{\input{fig/prop/14.tikz}}}\tabularnewline
 &  &  &  &  &  & \tabularnewline
\raisebox{0.5ex}{\scalebox{0.75}{\input{fig/prop/15.tikz}}} & \raisebox{0.5ex}{\scalebox{0.75}{\input{fig/prop/16.tikz}}} & \raisebox{0.5ex}{\scalebox{0.75}{\input{fig/prop/17.tikz}}} & \raisebox{0.5ex}{\scalebox{0.75}{\input{fig/prop/18.tikz}}} & \raisebox{0.5ex}{\scalebox{0.75}{\input{fig/prop/19.tikz}}} & \raisebox{0.5ex}{\scalebox{0.75}{\input{fig/prop/20.tikz}}} & \raisebox{0.5ex}{\scalebox{0.75}{\input{fig/prop/21.tikz}}}\tabularnewline
 &  &  &  &  &  & \tabularnewline
\raisebox{0.5ex}{\scalebox{0.75}{\input{fig/prop/22.tikz}}} & \raisebox{0.5ex}{\scalebox{0.75}{\input{fig/prop/23.tikz}}} & \raisebox{0.5ex}{\scalebox{0.75}{\input{fig/prop/24.tikz}}} & \raisebox{0.5ex}{\scalebox{0.75}{\input{fig/prop/25.tikz}}} & \raisebox{0.5ex}{\scalebox{0.75}{\input{fig/prop/26.tikz}}} & \raisebox{0.5ex}{\scalebox{0.75}{\input{fig/prop/27.tikz}}} & \raisebox{0.5ex}{\scalebox{0.75}{\input{fig/prop/28.tikz}}}\tabularnewline
\end{tabular}}
\par\end{centering}
\caption{\protect\label{tab:vvvv}Master integrals for 4-loop propagators.}
\end{table}

\subsection{Identifying integral families}

First let us start by identifying the integral families covering the
integrals in question. This is slightly more involved than with the
fully inclusive cuts for two reasons:
\begin{itemize}
\item there are more integral families because the $x$ variable can be
assigned to any of the cut lines, each assignment giving a different
integral;
\item because we want to treat the $\fn{\delta}{x-2q\sp p_{n}/q^{2}}$ factor
as a cut denominator (for IBP purposes), the denominator sets of many
diagrams turn out to be linearly dependent, so the integrands need
to be decomposed into partial fractions.
\end{itemize}
The procedure we have followed to identify the integral families is
as follows:\footnote{The code to perform each of these steps is available in
\noun{Alibrary} at \href{https://github.com/magv/alibrary}{github.com/magv/alibrary}.}
\begin{enumerate}
\item Use \noun{Qgraf}~\cite{Nog93} to generate all Feynman diagrams
of decays of off-shell photons, $Z$-bosons, and scalars that
couple to gluons into 2, 3, 4, and 5 particles up to 3 loops.
\item Insert Feynman rules into each diagram to obtain the amplitudes.
\item Construct all possible products of two amplitudes that are proportional
to $\alpha_{s}^{3}$.
\item Insert the $\fn{\delta}{x-2q\sp p_{n}/q^{2}}$ factor as an additional denominator. For uniformity with other denominators, we rewrite this factor as $q^2\, \fn{\delta}{(q-p)^2-(1-x)q^2}$.
\item Perform partial fraction decomposition for each product, transforming
each into one or several terms.
\item For each term of the partial fraction determine the set of denominators.
\item Pass the list of denominator sets to \noun{Feynson} (see \cite[Chap.\,4]{M22}),
and replace the loop momenta in each term as it suggests; making
symmetries explicit.
\item For each term compute its new denominator set, dropping any set that
is a subset of another. 
The resulting list of denominator sets are the needed unique integral families.
\end{enumerate}
In total we have found 256 integral families; they are summarized
in \ref{app:xcut-integral-families}: 18 with two cuts (\ref{tab:2px-cuts}),
34 with three cuts (\ref{tab:3px-cuts}), 96 with four cuts (\ref{tab:4px-cuts}),
and 108 with five cuts (\ref{tab:5px-cuts}). Each 2-particle cut
family has 11 denominators and an implied $\fn{\delta}{x-1}$; the
rest have 12 denominators, including $\fn{\delta}{x-2q\sp p_{n}/q^{2}}$.

\subsection{Identifying master integrals}

The next step is to determine the set of master integrals. We achieve 
this by first converting every term of the partial fractions to an
explicit $I_{\nu_{1},\dots,\nu_{n}}$ notation of \ref{eq:semiinclusive-int}, and passing the
list of these integrals to \noun{Fire6}~\cite{SC19} (used in combination with
\noun{LiteRed}~\cite{Lee14})---separately for each integral family, and for the
master integral list selection only. Alternatively we could have provided
any sufficiently large list of integrals to \noun{Fire6}; the only
advantage of our way is that we guarantee that no master integral
relevant to the decay cross-section computations was missed.

\noun{Fire6} can operate in two modes: the classical symbolic mode,
and the modular arithmetic mode, where all of the variables ($d$
and $x$) are replaced with fixed numerical values (e.g.\ 101921 and
961748927), and the reduction is performed modulo a large prime number
(e.g.\ 18446744073709551557): the latter is much faster (taking up
to several hours per integral family, where the symbolic mode is taking
up to several weeks), and generally produces the same set of master
integrals as the symbolic version. We have used three different combinations
of $d$, $x$, and the modulus values: all report the same set. In
the end up to 70 master integrals are identified per integral family.
After using \noun{Feynson} to identify symmetric integrals, the total
of 693 remain: 22 2-particle cuts (matching~\cite{HHM08,Hei+09,LSS10}),
96 3-particle cuts, 277 4-particle cuts, and 298
5-particle cuts. Note that this set is almost certainly overdetermined:
even though no two integrals are symmetric, some are linearly dependent,
because only removing symmetries is insufficient to undo differences
in the master integral selection per family. We do not see this as
a problem, however; having results for more integrals than necessary
simplifies, not complicates their usage.

\subsection{Constructing and solving differential equations}

In order to calculate the master integrals, we shall use differential
equations.
To write them down, for each integral family we differentiate
the master integrals by $x$ using
\begin{equation}
\frac{\partial}{\partial x}\fn{\delta}{x-2\frac{q\sp p_{n}}{q^{2}}}=-\frac{\fn{\delta}{x-2q\sp p_{n}/q^{2}}}{x-2q\sp p_{n}/q^{2}},
\end{equation}
express that in the $I_{\nu_{1},\dots,\nu_{n}}$ form, perform IBP
reduction for these derivatives, and construct the differential equation
matrix $\fn{\M M}{d,x}$:
\begin{equation}
    \partial_{x}\fn{\V I}{d,x}=\fn{\M M}{d,x}\fn{\V I}{d,x}.\label{eq:diff-eq}
\end{equation}
Overall, the reduction takes from several minutes up to 43 hours. In
the end, we obtain 256 matrices up to 70 \texttimes{} 70 in size, all
block-triangular (after sorting the master integrals by their propagator
set) with block sizes up to $4\times4$. Unfortunately, the differential
equation matrices that come directly out of this procedure sometimes
have inconvenient spurious poles poles like $1/\left(2x-1\right)$,
$d$-dependent poles like $1/\left(xd-2x-4d+19\right)$, poles given
by unfactorizable polynomials like $1/\left(x^{2}-3x-2\right)$, or
the same but also with $d$ dependence like $1/\left(3x^{2}d-14x^{2}+6xd-28x+d-6\right)$.
We see such artifacts only when there are more than one master integrals
per sector, and ideally we want none of them. For this reason we additionally
solve IBP relations for a small subset of integrals equal to the masters
but with raised indices; then, in each sector we can modify the master
integral selection (by choosing a subset of these) to try to get rid
of the inconvenient denominators. Fortunately, for block sizes up
to 4 trying even all possible combinations does not take too much
time. In all cases we are able to find a set of master integrals that
eliminate, or at least significantly reduce, the number of inconvenient
denominators in the diagonal blocks: all quadratic polynomials are
removed, and only rare spurious poles are left: in particular at $x=\pm2$,
but also in one case a pole of the form $1/\left(3+8\ep+2x-2\ep x\right)$
which is later removed during the $\ep$-form construction. The same
procedure also improves the off-diagonal blocks, even though those
are generally easier to handle, so we are less concerned about them.\footnote{%
For more disciplined methods of removing $d$-dependent poles see~\cite{Usovitsch20,SS20}.}

Next, we use \noun{Fuchsia} to construct such basis transformations,
\begin{equation}
    \fn{\V I}{d,x}=\fn{\M T}{d,x}\fn{\V J}{d,x},
    \label{eq:i-of-j}
\end{equation}
that reduce the differential equations to $\ep$-forms:
\begin{equation}
    \partial_{x}\fn{\V J}{d,x}=\ep\,\fn Sx\fn{\V J}{d,x}.
    \label{eq:epsilon-form}
\end{equation}
This takes from
minutes to days per integral family, with the most complex reduction
taking 14 days (the integral family in question is a 5-particle cut
family with 48 integrals, blocks up to $4\times4$, and the total
of 5~different poles, 2~of which are spurious, and the deepest being
$x^{-9}$). During the reduction spurious poles are necessarily removed,
and in the $\ep$-form the only remaining poles are located at
\begin{equation}
    x=\left\{ 0,+1,-1,+2,-2\right\} .\label{eq:alphabet}
\end{equation}
The last two poles are peculiar in that we expect them to be absent from physical amplitudes.
These poles prevent us from expressing integral values in terms of harmonic polylogarithms, which correspond to poles at $\left\{ 0,\pm1\right\}$.
Interestingly, they only remain in the off-diagonal blocks; even so, we have not found a way to completely remove them.
We have confirmed, however, that these poles cancel out during the calculation of the N$^3$LO semi-inclusive $e^+e^- \to q + X$ cross sections.
A similar behaviour was observed in the case of 3-loop deep inelastic scattering calculation~\cite{Vermaseren:2005qc}.

Next, we solve \ref{eq:epsilon-form} as 
\begin{equation}
    \begin{aligned}
        \fn{\M W}{d,x}&{}=\fn{\mathrm{P}\exp}{\ep\int_{x_{0}}^{x}\fn{\M S}y\d y}\equiv \\
        &\equiv\mathds{1}+\ep\int_{x_{0}}^{x}\d y_{1}\fn{\M S}{y_{1}}+\ep^{2}\int_{x_{0}}^{x}\d y_{1}\fn{\M S}{y_{1}}\int_{x_{0}}^{y_{1}}\d y_{2}\fn{\M S}{y_{2}}+\dots,    
    \end{aligned}
    \label{eq:fundamental-solution}
\end{equation}
\begin{equation}
    \fn{\V J}{d,x}=\fn{\M W}{d,x}\fn{\V C}d,
    \label{eq:j-of-fundamental-solution}
\end{equation}
where $\V C$ are the integration constants.

To determine $\V C$, it is sufficient to require that
an integral of $\fn I{d,x}$ over $x$ is equal to the corresponding
fully inclusive integral:
\begin{multline}
    \int_{0}^{1}\d x\undernote{\int\left(\prod_{i}\frac{\d^{d}l_{i}}{\left(2\pi\right)^{d}}\right)\fn{\d\PS_{n}}q\left(\prod_{i}\frac{1}{D_{i}^{\nu_{i}}}\right)\fn{\delta}{x-2\frac{q\sp p_{n}}{q^{2}}}}{\fn I{d,x}}=\\
    =\int\left.\left(\prod_{i}\frac{\d^{d}l_{i}}{\left(2\pi\right)^{d}}\right)\fn{\d\PS_{n}}q\left(\prod_{i}\frac{1}{D_{i}^{\nu_{i}}}\right)\right|_{x=2q\,\sp\,p_{n}/q^{2}}.
\end{multline}
Unfortunately, this is not completely straightforward in practice because
if we insert the series for $\fn I{d,x}$ into the left hand
side of this equation and integrate order by order, the integral will
likely diverge.

\subsubsection{Matching via integration: an illustration}

To demonstrate the matching process, let us take a look at the following 4-particle semi-inclusive cut of a 3-loop propagator:
\begin{equation}
    \fn I{d,x}=\raisebox{0.5ex}{\scalebox{0.75}{\input{fig/matchex.tikz}}}.
\end{equation}
The differential equation for it is
\begin{equation}
    \partial_{x}I=\left(\frac{1+2\ep}{1-x}-\frac{1+2\ep}{x}\right)I,
    \label{eq:exx-i-de}
\end{equation}
and an $\ep$-form can be achieved by changing the basis to such $J$,
that
\begin{equation}
\fn I{d,x}=\frac{1}{x\left(1-x\right)}\fn J{d,x}.\label{eq:exx-t}
\end{equation}
After the change of basis, the $\ep$-form equation reads
\begin{equation}
\partial_{x}J=\ep\left(\frac{2}{1-x}-\frac{2}{x}\right)J,\label{eq:exx-j-de}
\end{equation}
and the solution for it via \ref{eq:j-of-fundamental-solution} is
\begin{equation}
    \fn J{d,x}=\ep^{k_{0}}\left(C_{1}+\left(C_{2}-2C_{1}\fn G{0;x}-2C_{1}\fn G{1;x}\right)\ep+\fn{\mathcal{O}}{\ep^{2}}\right).\label{eq:exx-j-series}
\end{equation}
Here, $\fn{G}{w_1,\ldots,w_n;x}$ are Goncharov polylogarithms~\cite{Gon98}, or multiple polylogarithms.
The integration constants $C_{i}$ in principle can be determined
from the condition that
\begin{equation}
    \int\d x\,\fn I{d,x}\equiv\int\d x \, \raisebox{0.5ex}{\scalebox{0.75}{\input{fig/matchex.tikz}}}=\raisebox{0.5ex}{\scalebox{0.75}{\input{fig/matchex-inc.tikz}}}.\label{eq:exx-naive-condition}
\end{equation}
Because the transformation \ref{eq:exx-t} is singular at $x\to0$
and $x\to1$, an integral of $\fn I{d,x}$ defined that way will diverge
in each order of the series. Instead let us note that the same integral
but over $\fn J{d,x}$ will in fact converge, because as one can see
from \ref{eq:exx-j-series} at the boundaries of $x\to\left\{ 0,1\right\} $
$\fn J{d,x}$ diverges at most logarithmically (any multiple polylogarithm
does), and thus can be integrated over safely. This is a property
shared by all solutions to differential equations in an $\ep$-form---it
is only the basis change of \ref{eq:exx-t} that introduces the divergence.
Thus, to determine $C_{i}$ we can use
\begin{equation}
\int\d x\,\fn J{d,x}=\int\d x\,x\left(1-x\right)\raisebox{0.5ex}{\scalebox{0.75}{\input{fig/matchex.tikz}}}\overset{\text{IBP}}{=}-\frac{\ep}{1-4\ep}\raisebox{0.5ex}{\scalebox{0.75}{\input{fig/matchex-inc.tikz}}}.\label{eq:exx-improved-condition}
\end{equation}
Note that the right-hand side here is one order of $\ep$ lower than
it was in \ref{eq:exx-naive-condition}; this is because the prefactor
$x\left(1-x\right)$ has cancelled a divergence---exactly as we wanted.

Then, using the known value of
\begin{equation}
    \raisebox{0.5ex}{\scalebox{0.75}{\input{fig/matchex-inc.tikz}}}=
        \left(60\ep^{-4}-59\ep^{-3}+\fn{\mathcal{O}}{\ep^{-2}}\right)\raisebox{0.5ex}{\scalebox{0.75}{\input{fig/ps4.tikz}}},
    \label{eq:exx-inc-value}
\end{equation}
we can fix all the $C_{i}$.\footnote{Note that $\raisebox{0.5ex}{\scalebox{0.5}{\input{fig/matchex-inc.tikz}}}=\raisebox{0.5ex}{\scalebox{0.5}{\begin{tikzpicture}
	\begin{pgfonlayer}{nodelayer}
		\node [style=none] (0) at (-0.5, 0) {};
		\node [style=dot] (1) at (0, 0) {};
		\node [style=dot] (2) at (0.25, 0.75) {};
		\node [style=dot] (3) at (0.25, -0.75) {};
		\node [style=dot] (4) at (1, 0.75) {};
		\node [style=dot] (5) at (1, -0.75) {};
		\node [style=dot] (8) at (2, 0) {};
		\node [style=none] (9) at (2.5, 0) {};
		\node [style=dot] (6) at (1.25, 0) {};
	\end{pgfonlayer}
	\begin{pgfonlayer}{edgelayer}
		\draw [style=incoming edge] (0.center) to (1);
		\draw [style=outgoing edge] (8) to (9.center);
		\draw [style=cut edge] (2) to (4);
		\draw [style=cut edge] (2) to (5);
		\draw [style=cut edge] (3) to (4);
		\draw [style=cut edge] (3) to (5);
		\draw [style=edge, bend left=60] (6) to (8);
		\draw [style=edge, bend right=60] (6) to (8);
		\draw [style=edge] (1) to (2);
		\draw [style=edge] (1) to (3);
		\draw [style=edge] (4) to (6);
		\draw [style=edge] (5) to (6);
	\end{pgfonlayer}
\end{tikzpicture}
}}/\raisebox{0.5ex}{\scalebox{0.5}{\input{fig/bubble.tikz}}}$,
so we can reuse the results for 4-particle cuts of 4-loop propagators
to obtain 4-particle cuts of 3-loop propagators: all of the master
integrals of the latter are embedded in the former with just a bubble
attached.}
Inserting them into \ref{eq:exx-j-series}, the solution becomes
\begin{equation}
    \fn J{d,x}=\left(-60\ep^{-3}+\left(120\fn G{0;x}+120\fn G{1;x}+590\right)\ep^{-2}+\fn{\mathcal{O}}{\ep^{-1}}\right)\raisebox{0.5ex}{\scalebox{0.75}{\input{fig/ps4.tikz}}},\label{eq:exx-j-value}
\end{equation}
\begin{equation}
\fn I{d,x}=\left(\frac{-60}{x\left(1-x\right)}\ep^{-3}+\frac{120\fn G{0;x}+120\fn G{1;x}+590}{x\left(1-x\right)}\ep^{-2}+\fn{\mathcal{O}}{\ep^{-1}}\right)\raisebox{0.5ex}{\scalebox{0.75}{\input{fig/ps4.tikz}}}.\label{eq:exx-i-value}
\end{equation}

\subsubsection{The general case\protect\label{sec:matching-general-case}}

To calculate $C_{i}^{\left(k\right)}$ for the semi-inclusive master
integrals the procedure could be the same as \ref{eq:exx-improved-condition},
\begin{equation}
\int\d x\,\fn{\V J}{d,x}=\int_{0}^{1}\d x\,\fn{\M T^{-1}}{d,x}\fn{\V I}{d,x}=\left(\text{inclusive version of }\M T^{-1}\V I\right)\overset{\text{IBP}}{=}\dots,
\end{equation}
the only complication is performance: $\fn{\M T^{-1}}{d,x}$ might
contain $x$ to a high power, and because $x$ turns into a numerator
of the inclusive integral ($2q\sp p_{n}/q^{2}$), this results in longer IBP reduction times.
To avoid that and to minimize the numerator power we can instead use the condition
\begin{equation}
\int_{0}^{1}\d x\,x^{a}\left(1-x\right)^{b}\fn{\V I}{d,x}=\left(\text{inclusive version of }x^{a}\left(1-x\right)^{b}\V I\right)\overset{\text{IBP}}{=}\dots,\label{eq:matching-with-prefactors}
\end{equation}
where $a$ and $b$ are the smallest powers that compensate the divergence
of $\fn{\M T}{d,x}$ at $x\to0$ and $x\to1$, so that
\begin{equation}
x^{a}\fn{\M T}{d,x}=\mathcal{C}+\fn{\mathcal{O}}x\qquad\text{and}\qquad\left(1-x\right)^{b}\fn{\M T}{d,x}=\mathcal{C}'+\fn{\mathcal{O}}{1-x}.
\end{equation}
These conditions guarantee that the integrand in \ref{eq:matching-with-prefactors},
being $x^{a}\left(1-x\right)^{b}\M T\V J$, diverges at most logarithmically
at the limits (because of $\V J$), so the left-hand-side integral
converges, while the IBP reduction on the right-hand side needs to
deal with the smallest numerator powers.

We have performed this matching for each family of semi-inclusive
cuts of 4-loop propagators. Solving the IBP for inclusive integrals
on average took from an hour to several days, but several exceptional
cases required up to 6 weeks of time (this is with \noun{Fire6} running
on 16 cores for each family---because of the memory constraints we
could not afford to use more cores). The integration of the left-hand-side
of \ref{eq:matching-with-prefactors} also takes time: when $\fn{\V I}{d,x}$
is expanded via \ref{eq:fundamental-solution} up to 10~orders in
$\ep$ the integration takes from several minutes up to 4~days (using
our custom code, consisting of a mixture of \noun{Mathematica} and
\noun{Form}~\cite{RUTV17}), depending on the number of master integrals and the
number of poles they have. The number of terms in the ansatz for each
$\fn{\V I}{d,x}$ is roughly $\left(\text{number of poles}\right)^{\left(\text{orders of expansion}\right)}$,
which means that there is a practical cutoff on how many orders of
the expansion the matching conditions can be solved for.

A different problem is that because $x^{a}\left(1-x\right)^{b}\fn{\M T}{d,x}$
may have poles at values of $x$ other than $\left\{ 0,\pm1\right\} $,
for example at $x=\pm2$, the left-hand-side of \ref{eq:matching-with-prefactors}
results in multiple polylogarithms with corresponding poles in the
parameters (that is, $\pm2$). This is a complication because if the
parameters are restricted to only $\left\{ 0,\pm1\right\} $, then
all the integrals are expressible in terms of multiple zeta values (MZVs),
the relations between which are well known from e.g.~\cite{BBV10},
but with $\pm2$ in the parameter list, the constants that appear
are less studied. On the other hand, because these poles are only
an artifact of the way we match in \ref{eq:matching-with-prefactors},
and e.g.\ \ref{eq:exx-naive-condition} would be free from them, it
is expected that the values of $C_{i}^{\left(k\right)}$ should have
no traces of these. This is indeed the case, and for all $C_{i}$
that contain $\fn G{\dots,\pm2,\dots;1}$ we try to reduce them to
MZVs; to this end, we first evaluate them numerically with \noun{GiNaC}~\cite{VW05,BFK02},
and then use the PSLQ algorithm~\cite{FBA99} to restore them in the linear basis of MZVs.
This way we remove the artifacts of the matching procedure. On the other
hand, because the $\ep$-form differential equations themselves sometimes
contain $\pm2$ (see the list of poles in \ref{eq:alphabet}), the
$\fn G{\dots,\pm2,\dots;1}$ constants coming from that are not always
expressible in terms of MZVs. In these cases, we leave them as they
are without further reduction.

In the end we obtain $\ep$-series for all master integrals of semi-inclusive 2-, 3-, 4-, and 5-particle cuts of 4-loop propagators. 
We provide them in machine-readable form as supplementary material: see \ref{app:results} for a detailed description.

\subsection{Factorizing boundary behaviour\protect\label{subsec:factorizing-boundary-behavior}}

To use our integral values in practice we need to additionally know their boundary behaviour.
To see this, let us go back to the condition of \ref{eq:exx-naive-condition}.
Integrating over $\fn I{d,x}$ should give us $\raisebox{0.5ex}{\scalebox{0.5}{\input{fig/matchex-inc.tikz}}}$,
a well-defined quantity, and yet integration of $\fn I{d,x}$ defined
via \ref{eq:exx-t} and \ref{eq:exx-j-series} order-by-order in $\ep$
diverges. Why?

The divergence is a caused by the fact that the expansion in the
$\ep$-series here does not capture the full behaviour of $I$ at $x\to0$
and $x\to1$. This should not be surprising seeing that \ref{eq:exx-inc-value}
is one order higher in $\ep$ than \ref{eq:exx-i-value}, and this
order can never be recovered if one integrates order by order. Fortunately,
the behaviour of $\fn J{d,x}$ at $x\to0$ and $x\to1$ is fully determined
by the differential equation \ref{eq:exx-j-de}. To allow for the
correct integration, we only need to make this behaviour explicit.
To this end, let us expand \ref{eq:exx-j-de} in $x$ around the boundaries:
\begin{equation}
    \fn J{d,x\to0}\approx K_{0}x^{-2\ep},\qquad\text{and}\qquad\fn J{d,x\to1}\approx K_{1}\left(1-x\right)^{-2\ep}.\label{eq:exx-j-boundary-behavior}
\end{equation}
So to fix the integration problem, the form of $\fn J{d,x}$ should
become exactly this in the corresponding limits. In other words, we
should represent \ref{eq:exx-j-value} as
\begin{equation}
    \fn J{d,x}=x^{-2\ep}\left(1-x\right)^{-2\ep}\left(-60\ep^{-3}+590\ep^{-2}+\fn{\mathcal{O}}{\ep^{-1}}\right)\raisebox{0.5ex}{\scalebox{0.75}{\input{fig/ps4.tikz}}},
\end{equation}
and then $I$ accordingly becomes
\begin{equation}
    \fn I{d,x}=x^{-1-2\ep}\left(1-x\right)^{-1-2\ep}\left(-60\ep^{-3}+590\ep^{-2}+\fn{\mathcal{O}}{\ep^{-1}}\right)\raisebox{0.5ex}{\scalebox{0.75}{\input{fig/ps4.tikz}}}.\label{eq:exx-i-factored-value}
\end{equation}
This form can now be conveniently integrated order by order, if one
notes that
\begin{equation}
    \int_{0}^{1}\d x\,x^{-1-2\ep}\left(1-x\right)^{-1-2\ep}=\frac{\fn{\Gamma^{2}}{-2\ep}}{\fn{\Gamma}{-4\ep}}=-\frac{1}{\ep}+0+\fn{\mathcal{O}}{\ep}.\label{eq:exx-prefactor-int}
\end{equation}
Using this relation we can easily see that the integral of \ref{eq:exx-i-factored-value}
is indeed equal to \ref{eq:exx-inc-value}.

For more complicated integrals there are additional things to consider:
\begin{enumerate}
\item There can be a mix of different powers in the boundary expansion in
\ref{eq:exx-j-boundary-behavior}, both at $x\to0$ and at $x\to1$.
If this is so, then we shall present the integral as a sum of multiple
terms, each with a different prefactor being a combination of $x$
and $1-x$ powers.
\item Once the boundary behaviour is separated into an irrational prefactor,
the remaining series in $\ep$ in \ref{eq:exx-i-factored-value} need
not be just a constant: it can contain additional $x$ dependence.
In such cases integrating the obtained expressions is not as easy
as \ref{eq:exx-prefactor-int}. Let us see how this can be done.
\end{enumerate}

\subsubsection{The case of multiple irrational prefactors}

The general way to factor out the boundary behaviour is to start with
the $\ep$-form \ref{eq:epsilon-form}, and only consider its behaviour
at $x\to0$ and $x\to1$:
\begin{align}
    \partial_{x}\fn{\V J}{d,x} &= \ep\frac{\M S_{0}}{x}\left(1+\fn{\mathcal{O}}x\right)\fn{\V J}{d,x}, \\
    \partial_{x}\fn{\V J}{d,x} &= \ep\frac{\M S_{1}}{x-1}\left(1+\fn{\mathcal{O}}{x-1}\right)\fn{\V J}{d,x};
\end{align}
these equations can then be solved as
\begin{align}
    \fn{\V J}{d,x} &= e^{\ep\M S_{0}\ln x}\left(1+\fn{\mathcal{O}}x\right), \\
    \fn{\V J}{d,x} &= e^{\ep\M S_{1}\fn{\ln}{1-x}}\left(1+\fn{\mathcal{O}}{1-x}\right).
\end{align}
In the special case of the example of \ref{eq:exx-j-de} this gives
the results in \ref{eq:exx-j-boundary-behavior}. In more general
cases the matrix exponent needs to be calculated.\footnote{For example using the \code{MatrixExp} function in \noun{Mathematica}.}
Knowing this boundary behaviour we can then rewrite the $\ep$-form
solution given by \ref{eq:fundamental-solution} and \ref{eq:j-of-fundamental-solution}
as
\begin{equation}
    \fn{\V J}{d,x}=e^{\ep\M S_{1}\fn{\ln}{1-x}}\fn{\tilde{\M W}}{d,x}e^{\ep\M S_{0}\ln x}\,\fn{\V C}d,\label{eq:factorized-fundamental-solution}
\end{equation}
where $\fn{\tilde{\M W}}{d,x}$ is the fundamental solution regularized
at $x\to\left\{ 0,1\right\} $,
\begin{equation}
    \fn{\tilde{\M W}}{d,x}\equiv e^{-\ep\M S_{1}\fn{\ln}{1-x}}\fn{\tilde{\M W}}{d,x}e^{-\ep\M S_{0}\ln x}.
\end{equation}
This construction of $\fn{\tilde{\M W}}{d,x}$ ensures that no multiple
polylogarithm in its expansion has the form $\fn G{1,\dots;x}$ or
$\fn G{\dots,0;x}$, because all the leading ones and trailing zeros
have been factorized. The idea here is to only expand $\fn{\tilde{\M W}}{d,x}$
into a series in $\ep$, and keep the rest of the factors in \ref{eq:factorized-fundamental-solution}
unexpanded, just like in \ref{eq:exx-i-factored-value}.

\subsubsection{An example with multiple prefactors}

Let us illustrate the construction of \ref{eq:factorized-fundamental-solution}
with an example given by the integrals
\begin{equation}
\fn{\V I}{d,x}\equiv\left(\begin{array}{l}
I_{1}\\
I_{2}
\end{array}\right),\qquad\text{where}\qquad I_{i}\equiv\raisebox{0.5ex}{\scalebox{0.75}{\input{fig/ps4x.tikz}}},\qquad\text{and}\qquad I_{2}\equiv\raisebox{0.5ex}{\scalebox{0.75}{\input{fig/si/exx2.tikz}}}.
\end{equation}
The differential equation system for these two is
\begin{equation}
\partial_{x}\V I=\left(\begin{array}{cc}
\frac{1-2\ep}{x}+\frac{-1+2\ep}{1-x} & 0\\
\frac{-2+3\ep}{x}\frac{1}{q^{2}}+\frac{-2+3\ep}{1-x}\frac{1}{q^{2}} & \frac{1-3\ep}{x}
\end{array}\right)\V I,
\end{equation}
and the $\ep$-form can be achieved with the transformation found
by \noun{Fuchsia},
\begin{equation}
\V I=\left(\begin{array}{cc}
\ep x\left(1-x\right)q^{2} & 0\\
0 & \left(2-3\ep\right)x
\end{array}\right)\V J,\label{eq:exx2-t}
\end{equation}
giving us the $\ep$-form of
\begin{equation}
\partial_{x}\V J=\ep\left(\begin{array}{cc}
\frac{2}{1-x}-\frac{2}{x} & 0\\
-\frac{1}{x} & -\frac{3}{x}
\end{array}\right)\V J.
\end{equation}
The asymptotic behaviour of $\fn{\V J}{d,x}$ in the limit $x\to0$
can then be found as
\begin{equation}
\begin{aligned}
    \fn{\V J}{d,x}&=\fn{\exp}{\ep\left(\begin{array}{cc}
        -2 & 0\\
        -1 & -3
        \end{array}\right)\ln x}\left(\fn{\V C_{0}}d+\fn{\mathcal{O}}x\right)\\
    &=\left(\begin{array}{cc}
x^{-2\ep} & 0\\
x^{-3\ep}-x^{-2\ep} & x^{-3\ep}
\end{array}\right)\left(\fn{\V C_{0}}d+\fn{\mathcal{O}}x\right),
\end{aligned}
\end{equation}
where $\V C_{0}$ is a vector of integration constants.

Here we can see the main difference from the previous example: instead
of having a single irrational prefactor, $J_{2}$ (and thus, $I_{2}$)
behaves as a mixture of $x^{-2\ep}$ and $x^{-3\ep}$ for $x\to0$.

Similarly, for the $x\to1$ limit we have
\begin{equation}
\begin{aligned}
    \fn J{d,x}&=\fn{\exp}{\ep\left(\begin{array}{cc}
    -2 & 0\\
    0 & 0
    \end{array}\right)\fn{\ln}{1-x}}\left(\fn{\V C_{0}}d+\fn{\mathcal{O}}{1-x}\right)=\\
    &=\left(\begin{array}{cc}
\left(1-x\right)^{-2\ep} & 0\\
0 & 0
\end{array}\right)\left(\fn{\V C_{0}}d+\fn{\mathcal{O}}{1-x}\right).
\end{aligned}
\end{equation}
Taking into account the transformation from \ref{eq:exx2-t} we can
conclude that
\begin{equation}
I_{1}\propto x^{1-2x}\left(1-x\right)^{1-2\ep},
\end{equation}
which of course matches with what we expect from the full value of
$I_{1}$ given by \ref{eq:psnx}; $I_{2}$ on the other hand is a
mixture,
\begin{equation}
    I_{2}\propto Ax^{1-3\ep}+Bx^{1-2\ep}.
\end{equation}
Walking through the same steps of constructing the $\ep$-form solution
using \ref{eq:fundamental-solution}, matching with the fully inclusive
integrals, and applying \ref{eq:factorized-fundamental-solution},
we can obtain the following values of $\fn I{d,x}$:
\begin{align}
    I_{1}/\raisebox{0.5ex}{\scalebox{0.75}{\input{fig/ps4.tikz}}} & =x^{1-2\ep}\left(1-x\right)^{1-2\ep}\left(6-20\ep+\left(24\zeta_{2}-8\right)\ep^{2}+\dots\right),\nonumber \\
    I_{2}/\raisebox{0.5ex}{\scalebox{0.75}{\input{fig/ps4.tikz}}} & =x^{1-3\ep}\left(\frac{12}{\ep}-58+\left(72\zeta_{2}+44\right)\ep+\dots\right)+\label{eq:exx2-factorized-values}\\
 & +x^{1-2\ep}\left(-\frac{12}{\ep}+58+\left(24\fn G{0,1;x}-48\zeta_{2}-44\right)\ep+\dots\right).\nonumber 
\end{align}
The same procedure of factorizing the behaviour at boundaries can be
applied to the results for semi-inclusive cuts of 4-loop propagators.
We have done so, and have obtained the results in factorized form
similar to \ref{eq:exx2-factorized-values} for each master integral.
These results are available in machine-readable form, as described
in \ref{app:results}.

\subsection{Cross check}

As a check, we have calculated the N$^3$LO semi-inclusive cross-section for the $\gamma ~\rightarrow~ q + X$ process and integrated it over the $x$ parameter.
We have found the result to match the known fully inclusive results from~\cite{Inclusive_N3LO}. 

In general, these cross-sections contain terms of the form
\begin{equation}
    x^{-n_0+k_0\ep} \left(1-x\right)^{-n_1+k_1\ep} K(x),
\end{equation}
where $K(x)$ is regular in the $x \rightarrow 1$ and $x \rightarrow 0$ limits, and $n_0,n_1,k_0,k_1 \in \mathbb{Z}$.
Due to the divergences near $x=1$ and $x=0$, we can not integrate the expression order by order after series expansion in~$\ep$.
Instead, we use the following method:
\begin{equation}
    \begin{aligned}
    \int_{0}^{1} 
        x^{-n_0+k_0\ep}& \left(1-x\right)^{-n_1+k_1\ep} 
    K(x)\d x = \\
            &=
            \int_{0}^{1}
                x^{-n_0+k_0\ep}(1-x)^{-n_1+k_1\ep}
                \Bigg(
                K(x)
                -\sum_{i=0}^{n_0-1} x^{i}
                \frac{K^{i}(0)}{i!} \\
                &\qquad\qquad\qquad
                -\sum_{i=0}^{n_1-1} (-1)^{i}(x-1)^{i}
                \frac{K^{i}(1)}{i!}
                \Bigg)
            \d x
                \\
                &\qquad\qquad+
                \sum_{i=0}^{n_0-1}
                    \frac{1}{i+1+(-n_0+k_0\ep)}
                    \frac{K^{i}(0)}{i!}.
                \\
                &\qquad\qquad+
                \sum_{i=0}^{n_1-1}
                    \frac{(-1)^{i}}{i+1+(-n_1+k_1\ep)}
                    \frac{K^{i}(1)}{i!}.
    \end{aligned}
\end{equation}

\section{Semi-inclusive cuts of three-loop propagators}

The 3-loop semi-inclusive cut master integrals have been previously
computed in~\cite{Git16}.
As another cross-check we have re-computed these using the same
method as the 4-loop ones, and we confirm those results with the
following caveats:
\begin{enumerate}
\item The overall normalization factors are not listed in~\cite{Git16},
and are assumed to be the same as in~\cite{MM06}. The latter defines
its integrals in eq.~(A.1) and eq.~(A.11) of Appendix~A, with ``$V$''
integrals defined (in our notation) as
\begin{equation}
    \fn{V_{i_{1},\dots,i_{k}}}n\equiv\frac{e^{3\gamma_{E}\ep}}{\pi^{4}}\left(2\pi\right)^{3d-3}\int\frac{\d^{d}l}{\left(2\pi\right)^{d}}\d\PS_{3}\frac{\left(2q\sp p_{3}\right)^{n}}{D_{i_{1}}\cdots D_{i_{k}}},
\end{equation}
where $\d\PS_{n}$ is given by \ref{eq:dpsn}, and ``$R$'' integrals
as
\begin{equation}
    \fn{R_{i_{1},\dots,i_{k}}}n\equiv\frac{e^{3\gamma_{E}\ep}}{\pi^{3}}\left(2\pi\right)^{3d-4}\int\d\PS_{4}\frac{\left(2q\sp p_{4}\right)^{n}}{D_{i_{1}}\cdots D_{i_{k}}}.
    \label{eq:git16-r}
\end{equation}
However, to make the value of $\fn{R_{5}}0$ reported in eq.~(A.21)
consistent with \ref{eq:psn}, or indeed with that of~\cite{GGH04}
(which are the same), the prefactor in \ref{eq:git16-r} should read
$\frac{e^{3\gamma_{E}\ep}}{\pi^{3-3\ep}}$, not $\frac{e^{3\gamma_{E}\ep}}{\pi^{3}}$.
We have not been able to determine a similarly concise change for
the prefactors of $V$ integrals, and we can only report that the
values agree with ours up to a global constant factor.
\item Typos: in eq.~(A.21) \textquotedblleft $x^{-2\ep}$\textquotedblright{}
should read $\left(1-x\right)^{-2\ep}$, and \textquotedblleft $+2\zeta_{2}$\textquotedblright{}
should read $-2\zeta_{2}\mathrm{H}_{1}$. In eq.~(A.26) \textquotedblleft $\left(1-x\right)^{-1-2\ep}$\textquotedblright{}
should be ``$\left(1+x\right)^{-1-2\ep}$'' instead.\footnote{We have confirmed these errors with the author.}
\item The irrational prefactors separated in~\cite{Git16} are not not
equivalent to ours from \ref{subsec:factorizing-boundary-behavior},
because they don\textquoteright t always fully factorize the logarithms
of $x$ and $1-x$. Because of this we have only compared the series-expanded
forms of the results.
\end{enumerate}

\section{Summary\protect\label{chap:summary}}

We have presented the analytic calculation of all master integrals for 5-, 4-, and 3-particle semi-inclusive cuts of 4- and 3-loop massless propagators.
This completes the list of required master integrals for 3-loop perturbative corrections to semi-inclusive amplitudes of time-like annihilation processes.

As an immediate application, we see the extraction of photonic decay coefficient functions up to order~$\alpha_{s}^{3}$.
These are relevant to the analysis of semi-inclusive hadron production at $e^{+}e^{-}$ colliders (such as the planned FCC-ee, ILC, CLIC, and CEPC), and are currently only known up to $\alpha_{s}^{2}$.
Another application is the direct extraction of time-like splitting functions up to NNLO ($\alpha_{s}^3$) precision: together with the corresponding fragmentation functions these enter the analysis of semi-inclusive hadron production at both $e^{+}e^{-}$ and $pp$ colliders.
We leave these topics to follow-up publications~\cite{CFMM-to-appear}.

\section*{Acknowledgements}

We thank Sven-Olaf Moch, Oleksandr Gituliar, and Andrey Pikelner
for motivating this work.
The work of L.F. was supported by the German Academic Exchange Service (DAAD) through its Bi-Nationally Supervised Scholarship program.
This work has been supported by the Deutsche Forschungsgemeinschaft through the Research Unit FOR 2926, {\it Next Generation pQCD for Hadron Structure: Preparing for the EIC}, project number 40824754, and DFG grant MO~1801/4-2.

\section*{Note added}

While completing this work, we have been made aware of an independent work of \cite{other_N3LO_comp}, which encompasses the same calculation.

\appendix

\section{The resulting integral tables\protect\label{app:results}}

In the ancillary files of this article available at~\cite{xcut4lfiles} we provide the values of master integrals in both expanded and factorized forms.
The provided files are as follows:
\begin{description}
    \item[\textbf{\texttt{values-4l.m}}]\hfill\\
        A \noun{Mathematica} file listing integrals and their values,
        \begin{equation*}
            \texttt{\{\{} \text{integrand} \texttt{, } \text{value} \texttt{\},} \ldots\texttt{\}},
        \end{equation*}        
        where integrands are products of denominators defining the integral, and values are series' in $\ep$, as defined by \ref{eq:semiinclusive-int},
        and normalized as: 
        \begin{equation}
            \text{value}=I \, / \, \fn{\PS_{n}}{q} \, / \, \raisebox{0.5ex}{\scalebox{0.75}{\input{fig/bubble.tikz}}}^{m} ,
        \end{equation}
        where $n$ is the number of cut lines, and $m$ is the number of virtual loop integrations (i.e.\ $5-n$ for cuts of 4-loop propagators).
        The value of $\fn{\PS_{n}}{q}$ is given by \ref{eq:psn}, and
        \begin{equation}
            \raisebox{0.5ex}{\scalebox{0.75}{\input{fig/bubble.tikz}}}=\frac{i\pi^{\frac{d}{2}}}{\left(2\pi\right)^{d}}\frac{\fn{\Gamma^{2}}{\frac{d}{2}-1}\fn{\Gamma}{2-\frac{d}{2}}}{\fn{\Gamma}{d-2}}\left(-q^{2}-i0\right)^{\frac{d}{2}-2}.\label{eq:bubble-value}
        \end{equation}
        This normalization ensures that the listed values are dimensionless.

        The notation used in this file is:\vspace{-1ex}
        \begin{itemize}
            \item \texttt{ep}: dimensional regulator $\ep$;
            \item \texttt{q}: incoming momentum $q$;
            \item \texttt{l1}, \texttt{l2}, \texttt{l3}, \texttt{l4}: loop or external momenta being integrated over;
            \item \texttt{den[}$p$\texttt{]}: denominator, $1/p^2$;
            \item \texttt{cut[}$p$\texttt{]}: cut denominator, corresponding to an on-shell final-state particle, $2\pi \fn{\delta^{+}\!}{p^2}$;
            \item \texttt{xcut[}$p$\texttt{]}: $x$-tagged cut denominator, $q^2\fn{\delta}{p^2-(1-x)q^2}$;
            \item \texttt{G[}$w_1$\texttt{,}\ldots\texttt{,}$w_n$\texttt{,}$x$\texttt{]}: Goncharov (multiple) polylogarithm, $\fn{G}{w_1,\ldots,w_n;x}$;
            \item \texttt{Mzv[}$a$\texttt{,}$b$\texttt{,}\ldots\texttt{]}: multiple zeta value or an alternating sum, $\zeta_{a,b,\ldots}$, as defined in~\cite{BBV10}.
        \end{itemize}

    \item[\texttt{values-3l.m}]\hfill\\
        Same as above, but for 3-loop integrals.
    \item[\texttt{factored-values-4l.m}]\hfill\\
        A \noun{Mathematica} file listing integrals and their values with boundary behaviour factorised as described in \ref{subsec:factorizing-boundary-behavior},
        \begin{equation*}
            \texttt{\{\{} \text{integrand} \texttt{, \{} \text{prefactor} \texttt{, } \text{coefficient} \texttt{\},} \ldots\texttt{\},} \ldots\texttt{\}},
        \end{equation*}
        where prefactors contain the boundary behaviour,
        and coefficients are series' in $\ep$.
        This list is in the same order as \texttt{values-4l.m}.
    \item[\texttt{factored-values-3l.m}]\hfill\\
        Same as above, but for 3-loop integrals.
    \item[\texttt{denominators-4l.m}]\hfill\\
        A list of just the denominator products corresponding to integrals in \texttt{values-4l.m} and \texttt{factored-values-4l.m}.
    \item[\texttt{denominators-3l.m}]\hfill\\
        A list of just the denominator products corresponding to integrals in \texttt{values-3l.m} and \texttt{factored-values-3l.m}.
\end{description}

\section{Integral families for semi-inclusive cuts at four loops\protect\label{app:xcut-integral-families}}

We have identified the total of 256 integral families needed to cover
semi-inclusive decay at 4~loops. We list these families here: there
are 18 2-particle semi-inclusive cut families (\ref{tab:2px-cuts}),
34 3-cut families (\ref{tab:3px-cuts}), 96 4-cut families (\ref{tab:4px-cuts}),
and 108 5-cut particles (\ref{tab:5px-cuts}).

\begin{table}[h]
\begin{centering}
\scalebox{0.75}{%
\begin{tabular}{ccccccc}
$\raisebox{0.5ex}{\scalebox{0.75}{\input{fig/si/b1.tikz}}}$ & $\raisebox{0.5ex}{\scalebox{0.75}{\input{fig/si/b2.tikz}}}$ & $\raisebox{0.5ex}{\scalebox{0.75}{\input{fig/si/b3.tikz}}}$ & $\raisebox{0.5ex}{\scalebox{0.75}{\input{fig/si/b4.tikz}}}$ & $\raisebox{0.5ex}{\scalebox{0.75}{\input{fig/si/b5.tikz}}}$ & $\raisebox{0.5ex}{\scalebox{0.75}{\input{fig/si/b6.tikz}}}$ & $\raisebox{0.5ex}{\scalebox{0.75}{\input{fig/si/b7.tikz}}}$\tabularnewline
 &  &  &  &  &  & \tabularnewline
$\raisebox{0.5ex}{\scalebox{0.75}{\input{fig/si/b8.tikz}}}$ & $\raisebox{0.5ex}{\scalebox{0.75}{\input{fig/si/b9.tikz}}}$ & $\raisebox{0.5ex}{\scalebox{0.75}{\input{fig/si/b10.tikz}}}$ & $\raisebox{0.5ex}{\scalebox{0.75}{\input{fig/si/b11.tikz}}}$ & $\raisebox{0.5ex}{\scalebox{0.75}{\input{fig/si/b12.tikz}}}$ & $\raisebox{0.5ex}{\scalebox{0.75}{\input{fig/si/b13.tikz}}}$ & $\raisebox{0.5ex}{\scalebox{0.75}{\input{fig/si/b14.tikz}}}$\tabularnewline
 &  &  &  &  &  & \tabularnewline
$\raisebox{0.5ex}{\scalebox{0.75}{\input{fig/si/b15.tikz}}}$ & $\raisebox{0.5ex}{\scalebox{0.75}{\input{fig/si/b16.tikz}}}$ & $\raisebox{0.5ex}{\scalebox{0.75}{\input{fig/si/b17.tikz}}}$ & $\raisebox{0.5ex}{\scalebox{0.75}{\input{fig/si/b18.tikz}}}$ &  &  & \tabularnewline
 &  &  &  &  &  & \tabularnewline
\end{tabular}}
\par\end{centering}
\caption{\protect\label{tab:2px-cuts}2-particle-cut semi-inclusive families
at 4 loops.}
\end{table}

\begin{table}[h]
\begin{centering}
\scalebox{0.75}{%
\begin{tabular}{ccccccc}
$\raisebox{0.5ex}{\scalebox{0.75}{\input{fig/si/b21.tikz}}}$ & $\raisebox{0.5ex}{\scalebox{0.75}{\input{fig/si/b41.tikz}}}$ & $\raisebox{0.5ex}{\scalebox{0.75}{\input{fig/si/b43.tikz}}}$ & $\raisebox{0.5ex}{\scalebox{0.75}{\input{fig/si/b45.tikz}}}$ & $\raisebox{0.5ex}{\scalebox{0.75}{\input{fig/si/b46.tikz}}}$ & $\raisebox{0.5ex}{\scalebox{0.75}{\input{fig/si/b48.tikz}}}$ & $\raisebox{0.5ex}{\scalebox{0.75}{\input{fig/si/b52.tikz}}}$\tabularnewline
 &  &  &  &  &  & \tabularnewline
$\raisebox{0.5ex}{\scalebox{0.75}{\input{fig/si/b53.tikz}}}$ & $\raisebox{0.5ex}{\scalebox{0.75}{\input{fig/si/b55.tikz}}}$ & $\raisebox{0.5ex}{\scalebox{0.75}{\input{fig/si/b56.tikz}}}$ & $\raisebox{0.5ex}{\scalebox{0.75}{\input{fig/si/b60.tikz}}}$ & $\raisebox{0.5ex}{\scalebox{0.75}{\input{fig/si/b61.tikz}}}$ & $\raisebox{0.5ex}{\scalebox{0.75}{\input{fig/si/b62.tikz}}}$ & $\raisebox{0.5ex}{\scalebox{0.75}{\input{fig/si/b63.tikz}}}$\tabularnewline
 &  &  &  &  &  & \tabularnewline
$\raisebox{0.5ex}{\scalebox{0.75}{\input{fig/si/b64.tikz}}}$ & $\raisebox{0.5ex}{\scalebox{0.75}{\input{fig/si/b65.tikz}}}$ & $\raisebox{0.5ex}{\scalebox{0.75}{\input{fig/si/b67.tikz}}}$ & $\raisebox{0.5ex}{\scalebox{0.75}{\input{fig/si/b69.tikz}}}$ & $\raisebox{0.5ex}{\scalebox{0.75}{\input{fig/si/b70.tikz}}}$ & $\raisebox{0.5ex}{\scalebox{0.75}{\input{fig/si/b105.tikz}}}$ & $\raisebox{0.5ex}{\scalebox{0.75}{\input{fig/si/b106.tikz}}}$\tabularnewline
 &  &  &  &  &  & \tabularnewline
$\raisebox{0.5ex}{\scalebox{0.75}{\input{fig/si/b112.tikz}}}$ & $\raisebox{0.5ex}{\scalebox{0.75}{\input{fig/si/b113.tikz}}}$ & $\raisebox{0.5ex}{\scalebox{0.75}{\input{fig/si/b114.tikz}}}$ & $\raisebox{0.5ex}{\scalebox{0.75}{\input{fig/si/b115.tikz}}}$ & $\raisebox{0.5ex}{\scalebox{0.75}{\input{fig/si/b117.tikz}}}$ & $\raisebox{0.5ex}{\scalebox{0.75}{\input{fig/si/b127.tikz}}}$ & $\raisebox{0.5ex}{\scalebox{0.75}{\input{fig/si/b134.tikz}}}$\tabularnewline
 &  &  &  &  &  & \tabularnewline
$\raisebox{0.5ex}{\scalebox{0.75}{\input{fig/si/b135.tikz}}}$ & $\raisebox{0.5ex}{\scalebox{0.75}{\input{fig/si/b137.tikz}}}$ & $\raisebox{0.5ex}{\scalebox{0.75}{\input{fig/si/b138.tikz}}}$ & $\raisebox{0.5ex}{\scalebox{0.75}{\input{fig/si/b139.tikz}}}$ & $\raisebox{0.5ex}{\scalebox{0.75}{\input{fig/si/b169.tikz}}}$ & $\raisebox{0.5ex}{\scalebox{0.75}{\input{fig/si/b170.tikz}}}$ & \tabularnewline
 &  &  &  &  &  & \tabularnewline
\end{tabular}}
\par\end{centering}
\caption{\protect\label{tab:3px-cuts}3-particle-cut semi-inclusive families
at 4 loops.}
\end{table}

\begin{table}[p]
\begin{centering}
\scalebox{0.75}{%
\begin{tabular}{ccccccc}
$\raisebox{0.5ex}{\scalebox{0.75}{\input{fig/si/b19.tikz}}}$ & $\raisebox{0.5ex}{\scalebox{0.75}{\input{fig/si/b20.tikz}}}$ & $\raisebox{0.5ex}{\scalebox{0.75}{\input{fig/si/b22.tikz}}}$ & $\raisebox{0.5ex}{\scalebox{0.75}{\input{fig/si/b23.tikz}}}$ & $\raisebox{0.5ex}{\scalebox{0.75}{\input{fig/si/b25.tikz}}}$ & $\raisebox{0.5ex}{\scalebox{0.75}{\input{fig/si/b26.tikz}}}$ & $\raisebox{0.5ex}{\scalebox{0.75}{\input{fig/si/b27.tikz}}}$\tabularnewline
 &  &  &  &  &  & \tabularnewline
$\raisebox{0.5ex}{\scalebox{0.75}{\input{fig/si/b28.tikz}}}$ & $\raisebox{0.5ex}{\scalebox{0.75}{\input{fig/si/b29.tikz}}}$ & $\raisebox{0.5ex}{\scalebox{0.75}{\input{fig/si/b30.tikz}}}$ & $\raisebox{0.5ex}{\scalebox{0.75}{\input{fig/si/b31.tikz}}}$ & $\raisebox{0.5ex}{\scalebox{0.75}{\input{fig/si/b32.tikz}}}$ & $\raisebox{0.5ex}{\scalebox{0.75}{\input{fig/si/b36.tikz}}}$ & $\raisebox{0.5ex}{\scalebox{0.75}{\input{fig/si/b37.tikz}}}$\tabularnewline
 &  &  &  &  &  & \tabularnewline
$\raisebox{0.5ex}{\scalebox{0.75}{\input{fig/si/b38.tikz}}}$ & $\raisebox{0.5ex}{\scalebox{0.75}{\input{fig/si/b39.tikz}}}$ & $\raisebox{0.5ex}{\scalebox{0.75}{\input{fig/si/b50.tikz}}}$ & $\raisebox{0.5ex}{\scalebox{0.75}{\input{fig/si/b51.tikz}}}$ & $\raisebox{0.5ex}{\scalebox{0.75}{\input{fig/si/b54.tikz}}}$ & $\raisebox{0.5ex}{\scalebox{0.75}{\input{fig/si/b58.tikz}}}$ & $\raisebox{0.5ex}{\scalebox{0.75}{\input{fig/si/b59.tikz}}}$\tabularnewline
 &  &  &  &  &  & \tabularnewline
$\raisebox{0.5ex}{\scalebox{0.75}{\input{fig/si/b77.tikz}}}$ & $\raisebox{0.5ex}{\scalebox{0.75}{\input{fig/si/b78.tikz}}}$ & $\raisebox{0.5ex}{\scalebox{0.75}{\input{fig/si/b79.tikz}}}$ & $\raisebox{0.5ex}{\scalebox{0.75}{\input{fig/si/b80.tikz}}}$ & $\raisebox{0.5ex}{\scalebox{0.75}{\input{fig/si/b83.tikz}}}$ & $\raisebox{0.5ex}{\scalebox{0.75}{\input{fig/si/b84.tikz}}}$ & $\raisebox{0.5ex}{\scalebox{0.75}{\input{fig/si/b85.tikz}}}$\tabularnewline
 &  &  &  &  &  & \tabularnewline
$\raisebox{0.5ex}{\scalebox{0.75}{\input{fig/si/b86.tikz}}}$ & $\raisebox{0.5ex}{\scalebox{0.75}{\input{fig/si/b87.tikz}}}$ & $\raisebox{0.5ex}{\scalebox{0.75}{\input{fig/si/b91.tikz}}}$ & $\raisebox{0.5ex}{\scalebox{0.75}{\input{fig/si/b92.tikz}}}$ & $\raisebox{0.5ex}{\scalebox{0.75}{\input{fig/si/b93.tikz}}}$ & $\raisebox{0.5ex}{\scalebox{0.75}{\input{fig/si/b94.tikz}}}$ & $\raisebox{0.5ex}{\scalebox{0.75}{\input{fig/si/b95.tikz}}}$\tabularnewline
 &  &  &  &  &  & \tabularnewline
$\raisebox{0.5ex}{\scalebox{0.75}{\input{fig/si/b96.tikz}}}$ & $\raisebox{0.5ex}{\scalebox{0.75}{\input{fig/si/b97.tikz}}}$ & $\raisebox{0.5ex}{\scalebox{0.75}{\input{fig/si/b98.tikz}}}$ & $\raisebox{0.5ex}{\scalebox{0.75}{\input{fig/si/b99.tikz}}}$ & $\raisebox{0.5ex}{\scalebox{0.75}{\input{fig/si/b100.tikz}}}$ & $\raisebox{0.5ex}{\scalebox{0.75}{\input{fig/si/b101.tikz}}}$ & $\raisebox{0.5ex}{\scalebox{0.75}{\input{fig/si/b102.tikz}}}$\tabularnewline
 &  &  &  &  &  & \tabularnewline
$\raisebox{0.5ex}{\scalebox{0.75}{\input{fig/si/b103.tikz}}}$ & $\raisebox{0.5ex}{\scalebox{0.75}{\input{fig/si/b104.tikz}}}$ & $\raisebox{0.5ex}{\scalebox{0.75}{\input{fig/si/b107.tikz}}}$ & $\raisebox{0.5ex}{\scalebox{0.75}{\input{fig/si/b108.tikz}}}$ & $\raisebox{0.5ex}{\scalebox{0.75}{\input{fig/si/b109.tikz}}}$ & $\raisebox{0.5ex}{\scalebox{0.75}{\input{fig/si/b120.tikz}}}$ & $\raisebox{0.5ex}{\scalebox{0.75}{\input{fig/si/b121.tikz}}}$\tabularnewline
 &  &  &  &  &  & \tabularnewline
$\raisebox{0.5ex}{\scalebox{0.75}{\input{fig/si/b122.tikz}}}$ & $\raisebox{0.5ex}{\scalebox{0.75}{\input{fig/si/b123.tikz}}}$ & $\raisebox{0.5ex}{\scalebox{0.75}{\input{fig/si/b124.tikz}}}$ & $\raisebox{0.5ex}{\scalebox{0.75}{\input{fig/si/b144.tikz}}}$ & $\raisebox{0.5ex}{\scalebox{0.75}{\input{fig/si/b148.tikz}}}$ & $\raisebox{0.5ex}{\scalebox{0.75}{\input{fig/si/b151.tikz}}}$ & $\raisebox{0.5ex}{\scalebox{0.75}{\input{fig/si/b152.tikz}}}$\tabularnewline
 &  &  &  &  &  & \tabularnewline
$\raisebox{0.5ex}{\scalebox{0.75}{\input{fig/si/b153.tikz}}}$ & $\raisebox{0.5ex}{\scalebox{0.75}{\input{fig/si/b154.tikz}}}$ & $\raisebox{0.5ex}{\scalebox{0.75}{\input{fig/si/b155.tikz}}}$ & $\raisebox{0.5ex}{\scalebox{0.75}{\input{fig/si/b156.tikz}}}$ & $\raisebox{0.5ex}{\scalebox{0.75}{\input{fig/si/b157.tikz}}}$ & $\raisebox{0.5ex}{\scalebox{0.75}{\input{fig/si/b158.tikz}}}$ & $\raisebox{0.5ex}{\scalebox{0.75}{\input{fig/si/b159.tikz}}}$\tabularnewline
 &  &  &  &  &  & \tabularnewline
$\raisebox{0.5ex}{\scalebox{0.75}{\input{fig/si/b160.tikz}}}$ & $\raisebox{0.5ex}{\scalebox{0.75}{\input{fig/si/b161.tikz}}}$ & $\raisebox{0.5ex}{\scalebox{0.75}{\input{fig/si/b162.tikz}}}$ & $\raisebox{0.5ex}{\scalebox{0.75}{\input{fig/si/b163.tikz}}}$ & $\raisebox{0.5ex}{\scalebox{0.75}{\input{fig/si/b164.tikz}}}$ & $\raisebox{0.5ex}{\scalebox{0.75}{\input{fig/si/b165.tikz}}}$ & $\raisebox{0.5ex}{\scalebox{0.75}{\input{fig/si/b166.tikz}}}$\tabularnewline
 &  &  &  &  &  & \tabularnewline
$\raisebox{0.5ex}{\scalebox{0.75}{\input{fig/si/b167.tikz}}}$ & $\raisebox{0.5ex}{\scalebox{0.75}{\input{fig/si/b168.tikz}}}$ & $\raisebox{0.5ex}{\scalebox{0.75}{\input{fig/si/b171.tikz}}}$ & $\raisebox{0.5ex}{\scalebox{0.75}{\input{fig/si/b172.tikz}}}$ & $\raisebox{0.5ex}{\scalebox{0.75}{\input{fig/si/b173.tikz}}}$ & $\raisebox{0.5ex}{\scalebox{0.75}{\input{fig/si/b174.tikz}}}$ & $\raisebox{0.5ex}{\scalebox{0.75}{\input{fig/si/b175.tikz}}}$\tabularnewline
 &  &  &  &  &  & \tabularnewline
$\raisebox{0.5ex}{\scalebox{0.75}{\input{fig/si/b176.tikz}}}$ & $\raisebox{0.5ex}{\scalebox{0.75}{\input{fig/si/b177.tikz}}}$ & $\raisebox{0.5ex}{\scalebox{0.75}{\input{fig/si/b178.tikz}}}$ & $\raisebox{0.5ex}{\scalebox{0.75}{\input{fig/si/b179.tikz}}}$ & $\raisebox{0.5ex}{\scalebox{0.75}{\input{fig/si/b180.tikz}}}$ & $\raisebox{0.5ex}{\scalebox{0.75}{\input{fig/si/b181.tikz}}}$ & $\raisebox{0.5ex}{\scalebox{0.75}{\input{fig/si/b182.tikz}}}$\tabularnewline
 &  &  &  &  &  & \tabularnewline
$\raisebox{0.5ex}{\scalebox{0.75}{\input{fig/si/b183.tikz}}}$ & $\raisebox{0.5ex}{\scalebox{0.75}{\input{fig/si/b184.tikz}}}$ & $\raisebox{0.5ex}{\scalebox{0.75}{\input{fig/si/b185.tikz}}}$ & $\raisebox{0.5ex}{\scalebox{0.75}{\input{fig/si/b186.tikz}}}$ & $\raisebox{0.5ex}{\scalebox{0.75}{\input{fig/si/b194.tikz}}}$ & $\raisebox{0.5ex}{\scalebox{0.75}{\input{fig/si/b195.tikz}}}$ & $\raisebox{0.5ex}{\scalebox{0.75}{\input{fig/si/b196.tikz}}}$\tabularnewline
 &  &  &  &  &  & \tabularnewline
$\raisebox{0.5ex}{\scalebox{0.75}{\input{fig/si/b197.tikz}}}$ & $\raisebox{0.5ex}{\scalebox{0.75}{\input{fig/si/b198.tikz}}}$ & $\raisebox{0.5ex}{\scalebox{0.75}{\input{fig/si/b199.tikz}}}$ & $\raisebox{0.5ex}{\scalebox{0.75}{\input{fig/si/b200.tikz}}}$ & $\raisebox{0.5ex}{\scalebox{0.75}{\input{fig/si/b201.tikz}}}$ &  & \tabularnewline
 &  &  &  &  &  & \tabularnewline
\end{tabular}}
\par\end{centering}
\caption{\protect\label{tab:4px-cuts}4-particle-cut semi-inclusive families
at 4 loops.}
\end{table}

\begin{table}[p]
\begin{centering}
\scalebox{0.75}{%
\begin{tabular}{ccccccc}
$\raisebox{0.5ex}{\scalebox{0.75}{\input{fig/si/b24.tikz}}}$ & $\raisebox{0.5ex}{\scalebox{0.75}{\input{fig/si/b33.tikz}}}$ & $\raisebox{0.5ex}{\scalebox{0.75}{\input{fig/si/b34.tikz}}}$ & $\raisebox{0.5ex}{\scalebox{0.75}{\input{fig/si/b35.tikz}}}$ & $\raisebox{0.5ex}{\scalebox{0.75}{\input{fig/si/b40.tikz}}}$ & $\raisebox{0.5ex}{\scalebox{0.75}{\input{fig/si/b42.tikz}}}$ & $\raisebox{0.5ex}{\scalebox{0.75}{\input{fig/si/b44.tikz}}}$\tabularnewline
 &  &  &  &  &  & \tabularnewline
$\raisebox{0.5ex}{\scalebox{0.75}{\input{fig/si/b47.tikz}}}$ & $\raisebox{0.5ex}{\scalebox{0.75}{\input{fig/si/b49.tikz}}}$ & $\raisebox{0.5ex}{\scalebox{0.75}{\input{fig/si/b57.tikz}}}$ & $\raisebox{0.5ex}{\scalebox{0.75}{\input{fig/si/b66.tikz}}}$ & $\raisebox{0.5ex}{\scalebox{0.75}{\input{fig/si/b68.tikz}}}$ & $\raisebox{0.5ex}{\scalebox{0.75}{\input{fig/si/b71.tikz}}}$ & $\raisebox{0.5ex}{\scalebox{0.75}{\input{fig/si/b72.tikz}}}$\tabularnewline
 &  &  &  &  &  & \tabularnewline
$\raisebox{0.5ex}{\scalebox{0.75}{\input{fig/si/b73.tikz}}}$ & $\raisebox{0.5ex}{\scalebox{0.75}{\input{fig/si/b74.tikz}}}$ & $\raisebox{0.5ex}{\scalebox{0.75}{\input{fig/si/b75.tikz}}}$ & $\raisebox{0.5ex}{\scalebox{0.75}{\input{fig/si/b76.tikz}}}$ & $\raisebox{0.5ex}{\scalebox{0.75}{\input{fig/si/b81.tikz}}}$ & $\raisebox{0.5ex}{\scalebox{0.75}{\input{fig/si/b82.tikz}}}$ & $\raisebox{0.5ex}{\scalebox{0.75}{\input{fig/si/b88.tikz}}}$\tabularnewline
 &  &  &  &  &  & \tabularnewline
$\raisebox{0.5ex}{\scalebox{0.75}{\input{fig/si/b89.tikz}}}$ & $\raisebox{0.5ex}{\scalebox{0.75}{\input{fig/si/b90.tikz}}}$ & $\raisebox{0.5ex}{\scalebox{0.75}{\input{fig/si/b110.tikz}}}$ & $\raisebox{0.5ex}{\scalebox{0.75}{\input{fig/si/b111.tikz}}}$ & $\raisebox{0.5ex}{\scalebox{0.75}{\input{fig/si/b116.tikz}}}$ & $\raisebox{0.5ex}{\scalebox{0.75}{\input{fig/si/b118.tikz}}}$ & $\raisebox{0.5ex}{\scalebox{0.75}{\input{fig/si/b119.tikz}}}$\tabularnewline
 &  &  &  &  &  & \tabularnewline
$\raisebox{0.5ex}{\scalebox{0.75}{\input{fig/si/b125.tikz}}}$ & $\raisebox{0.5ex}{\scalebox{0.75}{\input{fig/si/b126.tikz}}}$ & $\raisebox{0.5ex}{\scalebox{0.75}{\input{fig/si/b128.tikz}}}$ & $\raisebox{0.5ex}{\scalebox{0.75}{\input{fig/si/b129.tikz}}}$ & $\raisebox{0.5ex}{\scalebox{0.75}{\input{fig/si/b130.tikz}}}$ & $\raisebox{0.5ex}{\scalebox{0.75}{\input{fig/si/b131.tikz}}}$ & $\raisebox{0.5ex}{\scalebox{0.75}{\input{fig/si/b132.tikz}}}$\tabularnewline
 &  &  &  &  &  & \tabularnewline
$\raisebox{0.5ex}{\scalebox{0.75}{\input{fig/si/b133.tikz}}}$ & $\raisebox{0.5ex}{\scalebox{0.75}{\input{fig/si/b136.tikz}}}$ & $\raisebox{0.5ex}{\scalebox{0.75}{\input{fig/si/b140.tikz}}}$ & $\raisebox{0.5ex}{\scalebox{0.75}{\input{fig/si/b141.tikz}}}$ & $\raisebox{0.5ex}{\scalebox{0.75}{\input{fig/si/b142.tikz}}}$ & $\raisebox{0.5ex}{\scalebox{0.75}{\input{fig/si/b143.tikz}}}$ & $\raisebox{0.5ex}{\scalebox{0.75}{\input{fig/si/b145.tikz}}}$\tabularnewline
 &  &  &  &  &  & \tabularnewline
$\raisebox{0.5ex}{\scalebox{0.75}{\input{fig/si/b146.tikz}}}$ & $\raisebox{0.5ex}{\scalebox{0.75}{\input{fig/si/b147.tikz}}}$ & $\raisebox{0.5ex}{\scalebox{0.75}{\input{fig/si/b149.tikz}}}$ & $\raisebox{0.5ex}{\scalebox{0.75}{\input{fig/si/b150.tikz}}}$ & $\raisebox{0.5ex}{\scalebox{0.75}{\input{fig/si/b187.tikz}}}$ & $\raisebox{0.5ex}{\scalebox{0.75}{\input{fig/si/b188.tikz}}}$ & $\raisebox{0.5ex}{\scalebox{0.75}{\input{fig/si/b189.tikz}}}$\tabularnewline
 &  &  &  &  &  & \tabularnewline
$\raisebox{0.5ex}{\scalebox{0.75}{\input{fig/si/b190.tikz}}}$ & $\raisebox{0.5ex}{\scalebox{0.75}{\input{fig/si/b191.tikz}}}$ & $\raisebox{0.5ex}{\scalebox{0.75}{\input{fig/si/b192.tikz}}}$ & $\raisebox{0.5ex}{\scalebox{0.75}{\input{fig/si/b193.tikz}}}$ & $\raisebox{0.5ex}{\scalebox{0.75}{\input{fig/si/b202.tikz}}}$ & $\raisebox{0.5ex}{\scalebox{0.75}{\input{fig/si/b203.tikz}}}$ & $\raisebox{0.5ex}{\scalebox{0.75}{\input{fig/si/b204.tikz}}}$\tabularnewline
 &  &  &  &  &  & \tabularnewline
$\raisebox{0.5ex}{\scalebox{0.75}{\input{fig/si/b205.tikz}}}$ & $\raisebox{0.5ex}{\scalebox{0.75}{\input{fig/si/b206.tikz}}}$ & $\raisebox{0.5ex}{\scalebox{0.75}{\input{fig/si/b207.tikz}}}$ & $\raisebox{0.5ex}{\scalebox{0.75}{\input{fig/si/b208.tikz}}}$ & $\raisebox{0.5ex}{\scalebox{0.75}{\input{fig/si/b209.tikz}}}$ & $\raisebox{0.5ex}{\scalebox{0.75}{\input{fig/si/b210.tikz}}}$ & $\raisebox{0.5ex}{\scalebox{0.75}{\input{fig/si/b211.tikz}}}$\tabularnewline
 &  &  &  &  &  & \tabularnewline
$\raisebox{0.5ex}{\scalebox{0.75}{\input{fig/si/b212.tikz}}}$ & $\raisebox{0.5ex}{\scalebox{0.75}{\input{fig/si/b213.tikz}}}$ & $\raisebox{0.5ex}{\scalebox{0.75}{\input{fig/si/b214.tikz}}}$ & $\raisebox{0.5ex}{\scalebox{0.75}{\input{fig/si/b215.tikz}}}$ & $\raisebox{0.5ex}{\scalebox{0.75}{\input{fig/si/b216.tikz}}}$ & $\raisebox{0.5ex}{\scalebox{0.75}{\input{fig/si/b217.tikz}}}$ & $\raisebox{0.5ex}{\scalebox{0.75}{\input{fig/si/b218.tikz}}}$\tabularnewline
 &  &  &  &  &  & \tabularnewline
$\raisebox{0.5ex}{\scalebox{0.75}{\input{fig/si/b219.tikz}}}$ & $\raisebox{0.5ex}{\scalebox{0.75}{\input{fig/si/b220.tikz}}}$ & $\raisebox{0.5ex}{\scalebox{0.75}{\input{fig/si/b221.tikz}}}$ & $\raisebox{0.5ex}{\scalebox{0.75}{\input{fig/si/b222.tikz}}}$ & $\raisebox{0.5ex}{\scalebox{0.75}{\input{fig/si/b223.tikz}}}$ & $\raisebox{0.5ex}{\scalebox{0.75}{\input{fig/si/b224.tikz}}}$ & $\raisebox{0.5ex}{\scalebox{0.75}{\input{fig/si/b225.tikz}}}$\tabularnewline
 &  &  &  &  &  & \tabularnewline
$\raisebox{0.5ex}{\scalebox{0.75}{\input{fig/si/b226.tikz}}}$ & $\raisebox{0.5ex}{\scalebox{0.75}{\input{fig/si/b227.tikz}}}$ & $\raisebox{0.5ex}{\scalebox{0.75}{\input{fig/si/b228.tikz}}}$ & $\raisebox{0.5ex}{\scalebox{0.75}{\input{fig/si/b229.tikz}}}$ & $\raisebox{0.5ex}{\scalebox{0.75}{\input{fig/si/b230.tikz}}}$ & $\raisebox{0.5ex}{\scalebox{0.75}{\input{fig/si/b231.tikz}}}$ & $\raisebox{0.5ex}{\scalebox{0.75}{\input{fig/si/b232.tikz}}}$\tabularnewline
 &  &  &  &  &  & \tabularnewline
$\raisebox{0.5ex}{\scalebox{0.75}{\input{fig/si/b233.tikz}}}$ & $\raisebox{0.5ex}{\scalebox{0.75}{\input{fig/si/b234.tikz}}}$ & $\raisebox{0.5ex}{\scalebox{0.75}{\input{fig/si/b235.tikz}}}$ & $\raisebox{0.5ex}{\scalebox{0.75}{\input{fig/si/b236.tikz}}}$ & $\raisebox{0.5ex}{\scalebox{0.75}{\input{fig/si/b237.tikz}}}$ & $\raisebox{0.5ex}{\scalebox{0.75}{\input{fig/si/b238.tikz}}}$ & $\raisebox{0.5ex}{\scalebox{0.75}{\input{fig/si/b239.tikz}}}$\tabularnewline
 &  &  &  &  &  & \tabularnewline
$\raisebox{0.5ex}{\scalebox{0.75}{\input{fig/si/b240.tikz}}}$ & $\raisebox{0.5ex}{\scalebox{0.75}{\input{fig/si/b241.tikz}}}$ & $\raisebox{0.5ex}{\scalebox{0.75}{\input{fig/si/b242.tikz}}}$ & $\raisebox{0.5ex}{\scalebox{0.75}{\input{fig/si/b243.tikz}}}$ & $\raisebox{0.5ex}{\scalebox{0.75}{\input{fig/si/b244.tikz}}}$ & $\raisebox{0.5ex}{\scalebox{0.75}{\input{fig/si/b245.tikz}}}$ & $\raisebox{0.5ex}{\scalebox{0.75}{\input{fig/si/b246.tikz}}}$\tabularnewline
 &  &  &  &  &  & \tabularnewline
$\raisebox{0.5ex}{\scalebox{0.75}{\input{fig/si/b247.tikz}}}$ & $\raisebox{0.5ex}{\scalebox{0.75}{\input{fig/si/b248.tikz}}}$ & $\raisebox{0.5ex}{\scalebox{0.75}{\input{fig/si/b249.tikz}}}$ & $\raisebox{0.5ex}{\scalebox{0.75}{\input{fig/si/b250.tikz}}}$ & $\raisebox{0.5ex}{\scalebox{0.75}{\input{fig/si/b251.tikz}}}$ & $\raisebox{0.5ex}{\scalebox{0.75}{\input{fig/si/b252.tikz}}}$ & $\raisebox{0.5ex}{\scalebox{0.75}{\input{fig/si/b253.tikz}}}$\tabularnewline
 &  &  &  &  &  & \tabularnewline
$\raisebox{0.5ex}{\scalebox{0.75}{\input{fig/si/b254.tikz}}}$ & $\raisebox{0.5ex}{\scalebox{0.75}{\input{fig/si/b255.tikz}}}$ & $\raisebox{0.5ex}{\scalebox{0.75}{\input{fig/si/b256.tikz}}}$ &  &  &  & \tabularnewline
 &  &  &  &  &  & \tabularnewline
\end{tabular}}
\par\end{centering}
\caption{\protect\label{tab:5px-cuts}5-particle-cut semi-inclusive families
at 4 loops.}
\end{table}

\section{Spherical coordinate system in \textit{n} dimensions\protect\label{app:spherical-coordinate-system}}

A spherical coordinate system in $n$ Euclidean dimensions parameterizes
a point $\vec{p}$ with a radius $r$, and angles $\theta_{1},\dots,\theta_{n-1}$,
such that the Cartesian coordinates corresponding to $\left\{ r,\theta\right\} $
are
\begin{equation}
\left(\vec{p}\right)_{i}=\begin{cases}
r\cos\theta_{1}, & \text{if }i=1,\\
r\sin\theta_{1}\cos\theta_{2}, & \text{if }i=2,\\
r\sin\theta_{1}\sin\theta_{2}\cos\theta_{3}, & \text{if }i=3,\\
\dots\\
r\sin\theta_{1}\sin\theta_{2}\cdots\cos\theta_{n-1}, & \text{if }i=n-1,\\
r\sin\theta_{1}\sin\theta_{2}\cdots\sin\theta_{n-1}, & \text{if }i=n,
\end{cases}
\end{equation}
where $r\ge0$, and $\theta_{i}$ are all assumed to lie in the range
of $\left[0;\pi\right]$, except for the last of them, $\theta_{n-1}$,
which lies in $\left[0;2\pi\right)$.

The volume element in spherical coordinates is
\begin{equation}
\d^{n}\vec{p}=\left|\det\frac{\partial\left\{ \vec{p}\right\} }{\partial\left\{ r,\theta\right\} }\right|\d\left\{ r,\theta\right\} =r^{n-1}\d r\undernote{\prod_{i=1}^{n-1}\sin^{n-1-i}\theta_{i}\,\d\theta_{i}}{=\d\Omega_{n-1}}.\label{eq:spherical-volume-element}
\end{equation}
An integral over the angular component of this volume element is the
total solid angle in $n$ dimensions, or the surface area of a unit
$(n-1)$-sphere. We shall denote this quantity as $\Omega_{n-1}$,
\begin{equation}
\Omega_{k}\equiv\int\prod_{i=1}^{k}\sin^{k-i}\theta_{i}\,\d\theta_{i}=\frac{2\pi^{\frac{k+1}{2}}}{\fn{\Gamma}{\frac{k+1}{2}}},\label{eq:omega}
\end{equation}
where we have used
\begin{equation}
\int_{0}^{\pi}\sin^{n}\theta\,\d\theta=\pi^{\frac{1}{2}}\frac{\fn{\Gamma}{\frac{n+1}{2}}}{\fn{\Gamma}{\frac{n+2}{2}}}.
\end{equation}

\bibliographystyle{JHEP}
\bibliography{main}

\providecommand{\href}[2]{#2}\begingroup\raggedright\begin{thebibliography}{10}

\bibitem{FCC18}
{\scshape FCC} collaboration, \emph{{FCC Physics Opportunities}: {Future
  Circular Collider Conceptual Design Report Volume 1}},
  \href{https://doi.org/10.1140/epjc/s10052-019-6904-3}{\emph{Eur. Phys. J. C}
  {\bfseries 79} (2019) 474}.

\bibitem{ILC13}
T.~Behnke, J.E.~Brau, B.~Foster, J.~Fuster, M.~Harrison, J.M.~Paterson,
  M.~Peskin, M.~Stanitzki, N.~Walker et~al., eds., \emph{{The International
  Linear Collider Technical Design Report - Volume 1: Executive Summary}},
  \href{https://arxiv.org/abs/1306.6327}{{\ttfamily 1306.6327}}.

\bibitem{CLIC18}
{\scshape CLICdp, CLIC} collaboration, \emph{{The Compact Linear Collider
  (CLIC) - 2018 Summary Report}},
  \href{https://arxiv.org/abs/1812.06018}{{\ttfamily 1812.06018}}.

\bibitem{CEPC2018}
{\scshape CEPC Study Group} collaboration, \emph{{CEPC Conceptual Design
  Report: Volume 2 - Physics \& Detector}},
  \href{https://arxiv.org/abs/1811.10545}{{\ttfamily 1811.10545}}.

\bibitem{RN96}
P.J.~Rijken and W.L.~van Neerven, \emph{{$O(\alpha_s^2)$ contributions to the
  longitudinal fragmentation function in $e^+ e^-$ annihilation}},
  \href{https://doi.org/10.1016/0370-2693(96)00898-2}{\emph{Phys. Lett. B}
  {\bfseries 386} (1996) 422}
  [\href{https://arxiv.org/abs/hep-ph/9604436}{{\ttfamily hep-ph/9604436}}].

\bibitem{RN97a}
P.J.~Rijken and W.L.~van Neerven, \emph{{Higher order QCD corrections to the
  transverse and longitudinal fragmentation functions in electron-positron
  annihilation}},
  \href{https://doi.org/10.1016/S0550-3213(96)00669-4}{\emph{Nucl. Phys. B}
  {\bfseries 487} (1997) 233}
  [\href{https://arxiv.org/abs/hep-ph/9609377}{{\ttfamily hep-ph/9609377}}].

\bibitem{RN97b}
P.J.~Rijken and W.L.~van Neerven, \emph{{$O(\alpha_s^2)$ contributions to the
  asymmetric fragmentation function in $e^+ e^-$ annihilation}},
  \href{https://doi.org/10.1016/S0370-2693(96)01529-8}{\emph{Phys. Lett. B}
  {\bfseries 392} (1997) 207}
  [\href{https://arxiv.org/abs/hep-ph/9609379}{{\ttfamily hep-ph/9609379}}].

\bibitem{MMV06}
A.~Mitov, S.~Moch and A.~Vogt, \emph{{Next-to-Next-to-Leading Order Evolution
  of Non-Singlet Fragmentation Functions}},
  \href{https://doi.org/10.1016/j.physletb.2006.05.005}{\emph{Phys. Lett. B}
  {\bfseries 638} (2006) 61}
  [\href{https://arxiv.org/abs/hep-ph/0604053}{{\ttfamily hep-ph/0604053}}].

\bibitem{MV08}
S.~Moch and A.~Vogt, \emph{{On third-order timelike splitting functions and
  top-mediated Higgs decay into hadrons}},
  \href{https://doi.org/10.1016/j.physletb.2007.10.069}{\emph{Phys. Lett. B}
  {\bfseries 659} (2008) 290}
  [\href{https://arxiv.org/abs/0709.3899}{{\ttfamily 0709.3899}}].

\bibitem{AMV12}
A.A.~Almasy, S.~Moch and A.~Vogt, \emph{{On the Next-to-Next-to-Leading Order
  Evolution of Flavour-Singlet Fragmentation Functions}},
  \href{https://doi.org/10.1016/j.nuclphysb.2011.08.028}{\emph{Nucl. Phys. B}
  {\bfseries 854} (2012) 133}
  [\href{https://arxiv.org/abs/1107.2263}{{\ttfamily 1107.2263}}].

\bibitem{Che+20}
H.~Chen, T.-Z.~Yang, H.X.~Zhu and Y.J.~Zhu, \emph{{Analytic Continuation and
  Reciprocity Relation for Collinear Splitting in QCD}},
  \href{https://doi.org/10.1088/1674-1137/abde2d}{\emph{Chin. Phys. C}
  {\bfseries 45} (2021) 043101}
  [\href{https://arxiv.org/abs/2006.10534}{{\ttfamily 2006.10534}}].

\bibitem{Git16}
O.~Gituliar, \emph{{Master integrals for splitting functions from differential
  equations in QCD}},
  \href{https://doi.org/10.1007/JHEP02(2016)017}{\emph{JHEP} {\bfseries 02}
  (2016) 017} [\href{https://arxiv.org/abs/1512.02045}{{\ttfamily
  1512.02045}}].

\bibitem{HHM08}
G.~Heinrich, T.~Huber and D.~{Ma\^itre}, \emph{{Master integrals for fermionic
  contributions to massless three-loop form-factors}},
  \href{https://doi.org/10.1016/j.physletb.2008.03.028}{\emph{Phys. Lett. B}
  {\bfseries 662} (2008) 344}
  [\href{https://arxiv.org/abs/0711.3590}{{\ttfamily 0711.3590}}].

\bibitem{Hei+09}
G.~Heinrich, T.~Huber, D.A.~Kosower and V.A.~Smirnov, \emph{{Nine-Propagator
  Master Integrals for Massless Three-Loop Form Factors}},
  \href{https://doi.org/10.1016/j.physletb.2009.06.038}{\emph{Phys. Lett. B}
  {\bfseries 678} (2009) 359}
  [\href{https://arxiv.org/abs/0902.3512}{{\ttfamily 0902.3512}}].

\bibitem{LSS10}
R.N.~Lee, A.V.~Smirnov and V.A.~Smirnov, \emph{{Analytic Results for Massless
  Three-Loop Form Factors}},
  \href{https://doi.org/10.1007/JHEP04(2010)020}{\emph{JHEP} {\bfseries 04}
  (2010) 020} [\href{https://arxiv.org/abs/1001.2887}{{\ttfamily 1001.2887}}].

\bibitem{M22}
V.~Magerya, \emph{{Semi- and Fully-Inclusive Phase-Space Integrals at Four
  Loops}}, Ph.D. thesis, Hamburg U., 2022.
\newblock \url{https://ediss.sub.uni-hamburg.de/bitstream/ediss/9787}.

\bibitem{xcut4lfiles}
V.~Magerya and L.~Fekésházy, \emph{{Values of master integrals for
  semi-inclusive cuts of massless three- and four-loop propagators}},  Mar.,
  2025.
\newblock
  \href{https://doi.org/10.5281/zenodo.15079434}{10.5281/zenodo.15079434}.

\bibitem{Hen13}
J.M.~Henn, \emph{{Multiloop integrals in dimensional regularization made
  simple}}, \href{https://doi.org/10.1103/PhysRevLett.110.251601}{\emph{Phys.
  Rev. Lett.} {\bfseries 110} (2013) 251601}
  [\href{https://arxiv.org/abs/1304.1806}{{\ttfamily 1304.1806}}].

\bibitem{Lee15}
R.N.~Lee, \emph{{Reducing differential equations for multiloop master
  integrals}}, \href{https://doi.org/10.1007/JHEP04(2015)108}{\emph{JHEP}
  {\bfseries 04} (2015) 108} [\href{https://arxiv.org/abs/1411.0911}{{\ttfamily
  1411.0911}}].

\bibitem{GM15}
O.~Gituliar and S.-O.~Moch, \emph{{Towards three-loop QCD corrections to the
  time-like splitting functions}},
  \href{https://doi.org/10.5506/APhysPolB.46.1279}{\emph{Acta Phys. Polon. B}
  {\bfseries 46} (2015) 1279}
  [\href{https://arxiv.org/abs/1505.02901}{{\ttfamily 1505.02901}}].

\bibitem{GM17}
O.~Gituliar and V.~Magerya, \emph{{\textsc{Fuchsia}: a tool for reducing
  differential equations for Feynman master integrals to epsilon form}},
  \href{https://doi.org/10.1016/j.cpc.2017.05.004}{\emph{Comput. Phys. Commun.}
  {\bfseries 219} (2017) 329}
  [\href{https://arxiv.org/abs/1701.04269}{{\ttfamily 1701.04269}}].

\bibitem{GM16}
O.~Gituliar and V.~Magerya, \emph{{\textsc{Fuchsia} and master integrals for
  splitting functions from differential equations in QCD}},
  \href{https://doi.org/10.22323/1.260.0030}{\emph{PoS} {\bfseries LL2016}
  (2016) 030} [\href{https://arxiv.org/abs/1607.00759}{{\ttfamily
  1607.00759}}].

\bibitem{MP19}
V.~Magerya and A.~Pikelner, \emph{{Cutting massless four-loop propagators}},
  \href{https://doi.org/10.1007/JHEP12(2019)026}{\emph{JHEP} {\bfseries 12}
  (2019) 026} [\href{https://arxiv.org/abs/1910.07522}{{\ttfamily
  1910.07522}}].

\bibitem{GMP18a}
O.~Gituliar, V.~Magerya and A.~Pikelner, \emph{{Five-Particle Phase-Space
  Integrals in QCD}},
  \href{https://doi.org/10.1007/JHEP06(2018)099}{\emph{JHEP} {\bfseries 06}
  (2018) 099} [\href{https://arxiv.org/abs/1803.09084}{{\ttfamily
  1803.09084}}].

\bibitem{MM06}
A.~Mitov and S.-O.~Moch, \emph{{QCD Corrections to Semi-Inclusive Hadron
  Production in Electron-Positron Annihilation at Two Loops}},
  \href{https://doi.org/10.1016/j.nuclphysb.2006.05.018}{\emph{Nucl. Phys. B}
  {\bfseries 751} (2006) 18}
  [\href{https://arxiv.org/abs/hep-ph/0604160}{{\ttfamily hep-ph/0604160}}].

\bibitem{Cza+15}
M.~Czakon, P.~Fiedler, T.~Huber, M.~Misiak, T.~Schutzmeier and M.~Steinhauser,
  \emph{{The $(Q_{7}, Q_{1,2})$ contribution to $ \overline{B}\to {X}_s\gamma $
  at $ \mathcal{O}\left({\alpha}_{\mathrm{s}}^2\right) $}},
  \href{https://doi.org/10.1007/JHEP04(2015)168}{\emph{JHEP} {\bfseries 04}
  (2015) 168} [\href{https://arxiv.org/abs/1503.01791}{{\ttfamily
  1503.01791}}].

\bibitem{BC10}
P.A.~Baikov and K.G.~Chetyrkin, \emph{{Four Loop Massless Propagators: An
  Algebraic Evaluation of All Master Integrals}},
  \href{https://doi.org/10.1016/j.nuclphysb.2010.05.004}{\emph{Nucl. Phys. B}
  {\bfseries 837} (2010) 186}
  [\href{https://arxiv.org/abs/1004.1153}{{\ttfamily 1004.1153}}].

\bibitem{LSS12}
R.N.~Lee, A.V.~Smirnov and V.A.~Smirnov, \emph{{Master Integrals for Four-Loop
  Massless Propagators up to Transcendentality Weight Twelve}},
  \href{https://doi.org/10.1016/j.nuclphysb.2011.11.005}{\emph{Nucl. Phys. B}
  {\bfseries 856} (2012) 95} [\href{https://arxiv.org/abs/1108.0732}{{\ttfamily
  1108.0732}}].

\bibitem{Nog93}
P.~Nogueira, \emph{{Automatic Feynman graph generation}},
  \href{https://doi.org/10.1006/jcph.1993.1074}{\emph{J. Comput. Phys.}
  {\bfseries 105} (1993) 279}.

\bibitem{SC19}
A.V.~Smirnov and F.S.~Chuharev, \emph{{\textsc{Fire6}: Feynman Integral
  REduction with Modular Arithmetic}},
  \href{https://doi.org/10.1016/j.cpc.2019.106877}{\emph{Comput. Phys. Commun.}
  {\bfseries 247 } (2020) 106877}
  [\href{https://arxiv.org/abs/1901.07808}{{\ttfamily 1901.07808}}].

\bibitem{Lee14}
R.N.~Lee, \emph{{\textsc{LiteRed} 1.4: a powerful tool for reduction of
  multiloop integrals}},
  \href{https://doi.org/10.1088/1742-6596/523/1/012059}{\emph{J. Phys. Conf.
  Ser.} {\bfseries 523} (2014) 012059}
  [\href{https://arxiv.org/abs/1310.1145}{{\ttfamily 1310.1145}}].

\bibitem{Usovitsch20}
J.~Usovitsch, \emph{{Factorization of denominators in integration-by-parts
  reductions}},  \href{https://arxiv.org/abs/2002.08173}{{\ttfamily
  2002.08173}}.

\bibitem{SS20}
A.V.~Smirnov and V.A.~Smirnov, \emph{{How to choose master integrals}},
  \href{https://doi.org/10.1016/j.nuclphysb.2020.115213}{\emph{Nucl. Phys. B}
  {\bfseries 960} (2020) 115213}
  [\href{https://arxiv.org/abs/2002.08042}{{\ttfamily 2002.08042}}].

\bibitem{Vermaseren:2005qc}
J.A.M.~Vermaseren, A.~Vogt and S.~Moch, \emph{{The Third-order QCD corrections
  to deep-inelastic scattering by photon exchange}},
  \href{https://doi.org/10.1016/j.nuclphysb.2005.06.020}{\emph{Nucl. Phys. B}
  {\bfseries 724} (2005) 3}
  [\href{https://arxiv.org/abs/hep-ph/0504242}{{\ttfamily hep-ph/0504242}}].

\bibitem{Gon98}
A.B.~Goncharov, \emph{{Multiple polylogarithms, cyclotomy and modular
  complexes}}, \href{https://doi.org/10.4310/MRL.1998.v5.n4.a7}{\emph{Math.
  Res. Lett.} {\bfseries 5} (1998) 497}
  [\href{https://arxiv.org/abs/1105.2076}{{\ttfamily 1105.2076}}].

\bibitem{RUTV17}
B.~Ruijl, T.~Ueda and J.~Vermaseren, \emph{{FORM version 4.2}},
  \href{https://arxiv.org/abs/1707.06453}{{\ttfamily 1707.06453}}.

\bibitem{BBV10}
{Bl\"umlein, J. and Broadhurst, D. J. and Vermaseren, J. A. M.}, \emph{{The
  Multiple Zeta Value Data Mine}},
  \href{https://doi.org/10.1016/j.cpc.2009.11.007}{\emph{Comput. Phys. Commun.}
  {\bfseries 181} (2010) 582}
  [\href{https://arxiv.org/abs/0907.2557}{{\ttfamily 0907.2557}}].

\bibitem{VW05}
J.~Vollinga and S.~Weinzierl, \emph{{Numerical evaluation of multiple
  polylogarithms}},
  \href{https://doi.org/10.1016/j.cpc.2004.12.009}{\emph{Comput. Phys. Commun.}
  {\bfseries 167} (2005) 177}
  [\href{https://arxiv.org/abs/hep-ph/0410259}{{\ttfamily hep-ph/0410259}}].

\bibitem{BFK02}
C.W.~Bauer, A.~Frink and R.~Kreckel, \emph{{Introduction to the GiNaC framework
  for symbolic computation within the C++ programming language}},
  \href{https://doi.org/10.1006/jsco.2001.0494}{\emph{J. Symb. Comput.}
  {\bfseries 33} (2002) 1} [\href{https://arxiv.org/abs/cs/0004015}{{\ttfamily
  cs/0004015}}].

\bibitem{FBA99}
H.~Ferguson, D.~Bailey and S.~Arno, \emph{{Analysis of PSLQ, an integer
  relation finding algorithm}},
  \href{https://doi.org/10.1090/S0025-5718-99-00995-3}{\emph{Math. Comp.}
  {\bfseries 68} (1999) 351}.

\bibitem{Inclusive_N3LO}
P.~Jakub\v{c}\'\i{}k, M.~Marcoli and G.~Stagnitto, \emph{{The parton-level
  structure of e$^{+}$e$^{-}$ to 2 jets at N$^{3}$LO}},
  \href{https://doi.org/10.1007/JHEP01(2023)168}{\emph{JHEP} {\bfseries 01}
  (2023) 168} [\href{https://arxiv.org/abs/2211.08446}{{\ttfamily
  2211.08446}}].

\bibitem{GGH04}
A.~Gehrmann-De~Ridder, T.~Gehrmann and G.~Heinrich, \emph{{Four particle phase
  space integrals in massless QCD}},
  \href{https://doi.org/10.1016/j.nuclphysb.2004.01.023}{\emph{Nucl. Phys. B}
  {\bfseries 682} (2004) 265}
  [\href{https://arxiv.org/abs/hep-ph/0311276}{{\ttfamily hep-ph/0311276}}].

\bibitem{CFMM-to-appear}
B.~Chargeishvili, L.~Fek\'esh\'azy, V.~Magerya and S.-O.~Moch, \emph{{To
  appear}},  2025.

\bibitem{other_N3LO_comp}
C.-Q.~He, H.~Xing, T.-Z.~Yang and H.X.~Zhu, \emph{{Single-inclusive hadron
  production in electron-positron annihilation at
  next-to-next-to-next-to-leading order in QCD}},  2025.
\newblock In preparation. Preprint number: ZU-TH~21/25.

\end{thebibliography}\endgroup

\end{document}